\documentstyle{mn}
\input{psfig.sty}
\newcommand{\etal}{{et al.~}}
\newcommand{\lta}{\la}
\newcommand{\gta}{\ga}

\newcommand{\kmsmpc}{\>{\rm km}\,{\rm s}^{-1}\,{\rm Mpc}^{-1}}
\newcommand{\kms}{\>{\rm km}\,{\rm s}^{-1}}

\newcommand{\Mpc}{\>{\rm Mpc}}

\newcommand{\Msun}{\>{\rm M_{\odot}}}

\newcommand{\Lsun}{\>{\rm L_{\odot}}}
\newcommand{\MLsun}{\>({\rm M}/{\rm L})_{\odot}}

\newcommand{\beq}{\begin{equation}}
\newcommand{\eeq}{\end{equation}}

\newcommand{\mpch}{\>h^{-1}{\rm {Mpc}}}

\newcommand{\msunh}{\>h^{-1}\rm M_\odot}
\newcommand{\Lsunhh}{\,h^{-2}\rm L_\odot}

\newcommand{\walpha}{\tilde{\alpha}}
\newcommand{\wLstar}{\tilde{L}^{*}}

\newcommand{\apj}{ApJ}
\newcommand{\apjs}{ApJS}
\newcommand{\aj}{AJ}
\newcommand{\mnras}{MNRAS}

\newcommand{\nat}{Nature}

\newdimen\hssize
\newdimen\hdsize 
\begin{document}


\title[Populating Dark Matter Haloes with galaxies]
      {Populating Dark Matter Haloes with Galaxies:
       Comparing the 2dFGRS with Mock Galaxy Redshift Surveys}
\author[Yang, Mo, Jing, van den Bosch \& Chu]
       {Xiaohu Yang$^{1,2}$, H.J. Mo$^{2,3}$, Y.P. Jing$^{4}$,
        Frank C. van den Bosch$^{3}$, YaoQuan Chu$^{1}$
        \thanks{E-mail: xhyang@ustc.edu.cn}\\
      $^1$Center for Astrophysics, University of Science and Technology
      of China, Hefei, Anhui 230026, China\\
      $^2$ Department of Astronomy, University of Massachusetts,
           Amherst MA 01003-9305, USA\\
      $^3$Max-Planck-Institut f\"ur Astrophysik
      Karl-Schwarzschild-Strasse 1, 85748 Garching, Germany\\
      $^4$Shanghai Astronomical Observatory; the Partner Group of MPA,
       Nandan Road 80,  Shanghai 200030, China}


\date{}


\maketitle

\label{firstpage}


\begin{abstract}
  In two recent papers, we  developed a powerful technique to link the
  distribution  of   galaxies  to  that  of  dark   matter  haloes  by
  considering halo occupation numbers as function of galaxy luminosity
  and  type.  In  this paper  we use  these distribution  functions to
  populate dark matter  haloes in high-resolution $N$-body simulations
  of   the  standard   $\Lambda$CDM  cosmogony   with  $\Omega_m=0.3$,
  $\Omega_{\Lambda}=0.7$,  and  $\sigma_8=0.9$.   Stacking  simulation
  boxes  of $100  h^{-1}  \Mpc$  and $300  h^{-1}  \Mpc$ with  $512^3$
  particles each  we construct Mock  Galaxy Redshift Surveys out  to a
  redshift  of $z=0.2$  with  a numerical  resolution that  guarantees
  completeness down  to $0.01  L^{*}$.  We use  these mock  surveys to
  investigate   various    clustering   statistics.    The   predicted
  two-dimensional  correlation function  $\xi(r_p,\pi)$  reveals clear
  signatures of redshift  space distortions. The projected correlation
  functions  for  galaxies  with  different  luminosities  and  types,
  derived from  $\xi(r_p,\pi)$, match the observations  well on scales
  larger than  $\sim 3\mpch$.  On  smaller scales, however,  the model
  overpredicts the  clustering power by about a  factor two.  Modeling
  the  ``finger-of-God''  effect  on  small scales  reveals  that  the
  standard $\Lambda$CDM  model predicts pairwise  velocity dispersions
  (PVD)  that  are  $\sim  400   \kms$  too  high  at  projected  pair
  separations  of $\sim  1 h^{-1}  \Mpc$.  A  strong velocity  bias in
  massive    haloes,   with    $b_{\rm    vel}   \equiv    \sigma_{\rm
  gal}/\sigma_{\rm  dm}  \sim   0.6$  (where  $\sigma_{\rm  gal}$  and
  $\sigma_{\rm dm}$ are the  velocity dispersions of galaxies and dark
  matter particles, respectively) can  reduce the predicted PVD to the
  observed level, but does not  help to resolve the over-prediction of
  clustering  power  on  small  scales.   Consistent  results  can  be
  obtained  within  the  standard  $\Lambda$CDM model  only  when  the
  average mass-to-light  ratio of  clusters is of  the order  of $1000
  \MLsun$  in the  $B$-band.  Alternatively,  as we  show by  a simple
  approximation, a $\Lambda$CDM model  with $\sigma_8 \simeq 0.75$ may
  also reproduce the observational results.  We discuss our results in
  light of the  recent WMAP results and the  constraints on $\sigma_8$
  obtained independently from other observations.
\end{abstract}


\begin{keywords}
dark matter  - large-scale structure of the universe - galaxies:
haloes - methods: statistical
\end{keywords}


\section{Introduction}

The distribution of galaxies  contains important information about the
large  scale structure of  the matter  distribution. On  large, linear
scales the galaxy power spectrum is believed to be proportional to the
matter   power  spectrum,   therewith  providing   useful  information
regarding  the  initial   conditions  of  structure  formation,  i.e.,
regarding the  power spectrum of primordial  density fluctuations.  On
smaller, non-linear scales the  distribution and motion of galaxies is
governed  by the  local  gravitational potential,  which is  cosmology
dependent. One of  the main goals of large  galaxy redshift surveys is
therefore  to  map  the  distribution  of galaxies  as  accurately  as
possible, over as  large a volume as possible.   The Sloan Digital Sky
Survey (SDSS; York \etal 2000)  and the 2 degree Field Galaxy Redshift
Survey   (2dFGRS;  Colless   \etal  2001)   are  two   of   the  prime
examples.  These surveys,  which are  currently being  completed, will
greatly enhance and improve our knowledge of large-scale structure and
will  become  the  standard  data  sets  against  which  to  test  our
cosmological and galaxy formation models for the decade to come.

However,  two effects complicate  a straightforward  interpretation of
the data. First  of all, the distribution of galaxies  is likely to be
biased with respect to the underlying mass density distribution.  This
bias is  an imprint of various complicated  physical processes related
to  galaxy formation  such as  gas cooling,  star  formation, merging,
tidal stripping and  heating, and a variety of  feedback processes. In
fact, it is expected that  the bias depends on scale, redshift, galaxy
type, galaxy  luminosity, etc.  (Kauffmann, Nusser  \& Steinmetz 1997;
Jing, Mo \& B\"orner 1998;  Somerville \etal 2001; van den Bosch, Yang
\& Mo 2003). Therefore, in  order to translate the observed clustering
of galaxies  into a measure for  the clustering of  (dark) matter, one
needs  to either  understand galaxy  formation  in detail,  or use  an
alternative method  to describe the relationship  between galaxies and
dark  matter (haloes).  One of  the  main goals  of this  paper is  to
advocate  one such  method  and  to show  its  potential strength  for
advancing our understanding of large scale structure.

Secondly,  because  of  the   peculiar  velocities  of  galaxies,  the
clustering of  galaxies observed  in redshift-space is  distorted with
respect  to the real-space  clustering (e.g.,  Davis \&  Peebles 1983;
Kaiser 1987; Regos \& Geller 1991; van de Weygaert \& van Kampen 1993;
Hamilton 1992).   On small scales,  the virialized motion  of galaxies
within dark matter haloes smears out structure along the line-of-sight
(i.e.,  the  so-called ``finger-of-God''  effect).   On large  scales,
coherent  flows induced  by the  gravitational action  of  large scale
structure  enhance structure  along the  line-of-sight.   Both effects
cause  an  anisotropy in  the  two-dimensional, two-point  correlation
function  $\xi(r_p,\pi)$, with  $r_p$ and  $\pi$ the  pair separations
perpendicular  and parallel  to the  line-of-sight,  respectively. The
large-scale flows compress the contours of $\xi(r_p,\pi)$ in the $\pi$
direction   by    an   amount   that   depends    on   $\beta   \equiv
\Omega_m^{0.6}/b$.   The  small-scale  peculiar motions  implies  that
$\xi(r_p,\pi)$ is convolved in the $\pi$-direction by the distribution
of pairwise velocities, $f(v_{12})$.   Thus, the detailed structure of
$\xi(r_p,\pi)$  contains information  regarding  the Universal  matter
density  $\Omega_m$,  the  (linear)  bias  of galaxies  $b$,  and  the
pairwise velocity distribution $f(v_{12})$.

From the above discussion it is obvious that understanding galaxy bias
is an integral  part of understanding large scale  structure.  One way
to address galaxy bias without a detailed theory for how galaxies form
is  to model  halo occupation  statistics. One  simply  specifies halo
occupation numbers,  $\langle N(M)  \rangle$, which describe  how many
galaxies  on  average  occupy  a   halo  of  mass  $M$.   Many  recent
investigations have used such  halo occupation models to study various
aspects of  galaxy clustering (Jing,  Mo \& B\"orner 1998;  Peacock \&
Smith  2000; Seljak 2000;  Scoccimarro \etal  2001; White  2001; Jing,
B\"orner \&  Suto 2002; Bullock, Wechsler \&  Somerville 2002; Berlind
\& Weinberg 2002;  Scranton 2002;  Kang \etal 2002; Marinoni \& Hudson
2002; Zheng \etal 2002; Kochanek \etal  2003).  In  two recent papers,
Yang, Mo \& van den Bosch (2003; hereafter Paper~I) and van den Bosch,
Yang \& Mo (2003; hereafter  Paper~II) have taken this halo occupation
approach one step further by  considering the occupation as a function
of  galaxy  luminosity  and  type.  They  introduced  the  conditional
luminosity function (hereafter CLF)  $\Phi(L \vert M) {\rm d}L$, which
gives the  number of  galaxies with luminosities  in the range  $L \pm
{\rm d}L/2$ that reside in haloes  of mass $M$.  The advantage of this
CLF over the  halo occupation function $\langle N(M)  \rangle$ is that
it allows one to address the clustering properties of galaxies {\it as
function of  luminosity}.  In addition,  the CLF yields a  direct link
between the halo mass function and the galaxy luminosity function, and
allows  a straightforward  computation  of the  average luminosity  of
galaxies residing in  a halo of given mass.   Therefore, $\Phi(L \vert
M)$ is not only constrained  by the clustering properties of galaxies,
as is the  case with $\langle N(M) \rangle$, but  also by the observed
LFs and the halo mass-to-light ratios.

In Papers~I  and~II we used the  observed LFs and  the luminosity- and
type-dependence  of  the  galaxy  two-point  correlation  function  to
constrain  the CLF in  the standard  $\Lambda$CDM cosmology.   In this
paper,  we   use  this   CLF  to  populate   dark  matter   haloes  in
high-resolution  $N$-body simulations.   The `virtual  Universes' thus
obtained  are used  to  construct mock  galaxy  redshift surveys  with
volumes and apparent magnitude limits  similar to those in the 2dFGRS.
This is  the first  time that realistic  mock surveys  are constructed
that  (i)  associate  galaxies  with  dark  matter  haloes,  (ii)  are
independent of a model for  how galaxies form, and (iii) automatically
have the correct galaxy abundances and correlation lengths as function
of  galaxy luminosity  and type.  In  the past,  mock galaxy  redshift
surveys  were constructed  either  by associating  galaxies with  dark
matter particles (rather than haloes)  using a completely {\it ad hoc}
bias scheme  (Cole \etal 1998),  or by linking  semi-analytical models
for galaxy formation (with  all their associated uncertainties) to the
merger  histories  of  dark   matter  haloes  derived  from  numerical
simulations (Kauffmann \etal 1999; Mathis \etal 2002).

We  use our mock  galaxy redshift  survey to  investigate a  number of
statistical measures of the  large scale distribution of galaxies.  In
particular, we focus on the two-point correlation function in redshift
space,  its distortions  on small  and  large scales,  and the  galaxy
pairwise   peculiar  velocities.   Where   possible  we   compare  our
predictions with  the 2dFGRS and  we discuss the sensitivity  of these
clustering statistics to several details regarding the halo occupation
statistics.  We  show that  the halo occupation  obtained analytically
can reliably be implemented in  $N$-body simulations. We find that the
standard $\Lambda$CDM model, together with the halo occupation we have
obtained, can reproduce many of the observational results. However, we
find  significant  discrepancy   between  the  model  predictions  and
observations on small scales.  We  show that to get consistent results
on  small scales,  either  the mass-to-light  ratios  for clusters  of
galaxies are significantly higher than normally assumed, or the linear
power spectrum has an amplitude  that is significantly lower than its
`concordance' value.

This paper is organized as follows. In Section~\ref{sec:clf} we review
the     CLF     formalism     developed    in     papers~I     and~II.
Section~\ref{sec:method}  introduces   the  $N$-body  simulations  and
describes  our  method  of  populating  dark matter  haloes  in  these
simulations   with  galaxies   of  different   type   and  luminosity.
Section~\ref{sec:resreal}  investigates several  clustering statistics
in  real-space and  focuses on  the  accuracy with  which mock  galaxy
distributions  can  be  constructed   using  our  CLF  formalism.   In
Section~\ref{sec:red}  we  use  these  mock  galaxy  distributions  to
construct  mock galaxy redshift  surveys that  are comparable  in size
with the  2dFGRS. We extract the  redshift-space two-point correlation
function from this mock  redshift survey, investigate its anisotropies
induced by  the galaxy  peculiar motions, and  compare our  results to
those  obtained   from  the  2dFGRS  by  Hawkins   \etal  (2003).   In
section~\ref{sec:disc}  we  discuss  possible  ways to  alleviate  the
discrepancy  between model and  observations on  small scales,  and we
summarize our results in Section~\ref{sec:concl}.

\section{The Conditional Luminosity Function}
\label{sec:clf}

In  Paper~I  we  developed  a  formalism,  based  on  the  conditional
luminosity  function $\Phi(L \vert  M)$, to  link the  distribution of
galaxies to that of dark matter haloes.  We introduced a parameterized
form for $\Phi(L  \vert M)$ which we constrained using  the LF and the
correlation  lengths  as  function  of  luminosity.   In  Paper~II  we
extended this  formalism by constructing separate CLFs  for the early-
and  late-type  galaxies.  In  this  paper  we  use these  results  to
populate   dark   matter  haloes,   obtained   from  large   numerical
simulations,  with both  early-  and late-type  galaxies of  different
luminosities.  For  completeness, we  briefly summarize here  the main
ingredients of  the CLF  formalism, and refer  the reader  to papers~I
and~II for more details.

The conditional luminosity  function is  parameterized by a  Schechter
function:
\begin{equation}
\label{phiLM}
\Phi(L  \vert  M)  {\rm  d}L  = {\tilde{\Phi}^{*}  \over  \wLstar}  \,
\left({L \over  \wLstar}\right)^{\walpha} \, \,  {\rm exp}(-L/\wLstar)
\, {\rm d}L,
\end{equation}
where   $\wLstar   =   \wLstar(M)$,   $\walpha   =   \walpha(M)$   and
$\tilde{\Phi}^{*}  = \tilde{\Phi}^{*}(M)$  are all  functions  of halo
mass  $M$\footnote{Halo masses are  defined as  the masses  within the
radius $r_{180}$  inside of which the average  overdensity is $180$.}.
Following Papers~I  and~II, we write the  average total mass-to-light
ratio of a halo of mass $M$ as
\begin{equation}
\label{MtoLmodel}
\left\langle {M \over L} \right\rangle(M) = {1 \over 2} \,
\left({M \over L}\right)_0 \left[ \left({M \over M_1}\right)^{-\gamma_1} +
\left({M \over M_1}\right)^{\gamma_2}\right],
\end{equation}
which has four free parameters: a characteristic mass $M_1$, for which
the  mass-to-light  ratio  is  equal  to $(M/L)_0$,  and  two  slopes,
$\gamma_1$ and  $\gamma_2$, that specify the behavior  of $\langle M/L
\rangle$  at the  low and  high  mass ends,  respectively.  A  similar
parameterization   is  adopted   for  the   characteristic  luminosity
$\wLstar(M)$:
\begin{equation}
\label{LstarM}
{M \over \wLstar(M)} = {1 \over 2} \, \left({M \over L}\right)_0 \,
f(\walpha) \, \left[ \left({M \over M_1}\right)^{-\gamma_1} +
\left({M \over M_2}\right)^{\gamma_3}\right],
\end{equation}
with
\begin{equation}
\label{falpha}
f(\walpha) = {\Gamma(\walpha+2) \over \Gamma(\walpha+1,1)}.
\end{equation}
Here  $\Gamma(x)$   is  the  Gamma  function   and  $\Gamma(a,x)$  the
incomplete Gamma  function.  This parameterization  has two additional
free  parameters: a characteristic  mass $M_2$  and a  power-law slope
$\gamma_3$.   For $\walpha(M)$ we  adopt a  simple linear  function of
$\log(M)$,
\begin{equation}
\label{alphaM}
\walpha(M) = \alpha_{15} + \eta \, \log(M_{15}),
\end{equation}
with $M_{15}$ the halo mass in units of $10^{15} \msunh$, $\alpha_{15}
= \walpha(M_{15}=1)$, and $\eta$ describes the change of the faint-end
slope  $\walpha$ with  halo  mass.  Note  that  once $\walpha(M)$  and
$\wLstar  (M)$ are  given, the  normalization of  the  conditional LF,
$\tilde{\Phi}^{*}(M)$,  is  obtained  through  equations~(\ref{phiLM})
and~(\ref{MtoLmodel}),  using  the   fact  that  the  total  (average)
luminosity in a halo of mass $M$ is
\begin{equation}
\label{meanL}
\langle L \rangle(M) = \int_{0}^{\infty}  \Phi(L \vert M) \, L
\, {\rm d}L = \tilde{\Phi}^{*} \, \wLstar \, \Gamma(\walpha+2).
\end{equation}
Finally, we introduce the mass  scale $M_{\rm min}$ below which we set
the CLF to zero; i.e., we assume that no stars form inside haloes with
$M  < M_{\rm  min}$.   Motivated by  reionization considerations  (see
Paper~I  for details)  we adopt  $M_{\rm min}  = 10^{9}  h^{-1} \Msun$
throughout.

In order  to split the galaxy  population in early and  late types, we
follow Paper~II and introduce  the function $f_{\rm late}(L,M)$, which
specifies the  fraction of galaxies  with luminosity $L$ in  haloes of
mass  $M$  that are  late-type.   The  CLFs  of late-  and  early-type
galaxies are then given by
\begin{equation}
\label{CLFl}
\Phi_{\rm late}(L \vert M) {\rm d}L = f_{\rm late}(L,M) \, \Phi(L
\vert M) {\rm d}L
\end{equation}
and
\begin{equation}
\label{CLFe}
\Phi_{\rm early}(L \vert M) \, {\rm d}L = \left[ 1 - f_{\rm late}(L,M)
\right] \, \Phi(L \vert M) \, {\rm d}L\,.
\end{equation}
As  with the  CLF for  the entire  population of  galaxies, $\Phi_{\rm
late}(L \vert M)$ and $\Phi_{\rm early}(L \vert M)$ are constrained by
2dFGRS measurements of the LFs and the correlation lengths as function
of   luminosity.    We  assume   that   $f_{\rm   late}(L,M)$  has   a
quasi-separable form
\begin{equation}
\label{fracdef}
f_{\rm late}(L,M) = g(L) \, h(M) \, q(L,M).
\end{equation}
Here
\begin{equation}
\label{qlm}
q(L,M) = \left\{
\begin{array}{lll}
1                      & \mbox{if $g(L) \, h(M) \leq 1$} \\
{1 \over g(L) \, h(M)} & \mbox{if $g(L) \, h(M) > 1$}
\end{array} \right.
\end{equation}
is to ensure that $f_{\rm late}(L,M) \leq 1$. We adopt
\begin{equation}
\label{gl}
g(L) = {\hat{\Phi}_{\rm late}(L) \over \hat{\Phi}(L)}
{\int_{0}^{\infty} \Phi(L \vert M) \, n(M) \, {\rm d}M \over
 \int_{0}^{\infty} \Phi(L \vert M) \, h(M) \, n(M) \, {\rm d}M}
\end{equation}
where $n(M)$ is the halo mass function 
(Sheth \& Tormen 1999; Sheth, Mo \& Tormen 2001),
 $\hat{\Phi}_{\rm late}(L)$ and $\hat{\Phi}(L)$ correspond to the
{\it  observed}  LFs  of  the  late-type and  entire  galaxy  samples,
respectively, and

\begin{equation}
\label{hm}
h(M) = \max \left( 0, \min\left[ 1, \left({{\rm log}(M/M_a)
\over {\rm log}(M_b/M_a)} \right) \right] \right)
\end{equation}
with $M_a$  and $M_b$ two  additional free parameters, defined  as the
masses at which $h(M)$ takes  on the values $0$ and $1$, respectively.
As shown  in Paper~II, this parameterization allows  the population of
galaxies  to  be  split  in  early- and  late-types  such  that  their
respective LFs and clustering properties are well fitted.

In  Papers~I  and~II we  presented  a  number  of different  CLFs  for
different  cosmologies and  different assumptions  regarding  the free
parameters.   In  what  follows  we  focus on  the  flat  $\Lambda$CDM
cosmology with  $\Omega_m=0.3$, $\Omega_{\Lambda}=0.7$ and $h=H_0/(100
\kmsmpc) = 0.7$  and with initial density fluctuations  described by a
scale-invariant  power  spectrum  with normalization  $\sigma_8=0.9$.  
These cosmological parameters are in  good agreement with a wide range
of  observations, including  the  recent WMAP  results (Spergel  \etal
2003),  and in  what follows  we refer  to it  as  the ``concordance''
cosmology.  Finally,  we adopt the CLF with  the following parameters:
$M_1  =  10^{10.94}  h^{-1}  \Msun$,  $M_2=10^{12.04}  h^{-1}  \Msun$,
$M_a=10^{17.26}   h^{-1}   \Msun$,   $M_b=10^{10.86}  h^{-1}   \Msun$,
$(M/L)_0=124    h     \MLsun$,    $\gamma_1=2.02$,    $\gamma_2=0.30$,
$\gamma_3=0.72$,  $\eta=-0.22$  and  $\alpha_{15}=-1.10$.  This  model
(referred  to as  model~D in  Paper~II) yields  excellent fits  to the
observed LFs and the observed  correlation lengths as function of both
luminosity and type\footnote{Note that  the parameters listed here are
  slightly  different  from those  given  in  the  orignal version  of
  Paper~II, as they are based on a corrected version of the galaxy
  luminosity function.
  As shown in Paper I, a change in the overall 
  amplitude of the luminosity function in the fitting has some effect 
  on the best-fit values of the correlation lengths.
  This is due to the combination of the following two effects.
  First, our model assumes a fixed mass-to-light 
  ratio for massive haloes and so a change in the amplitude of the 
  luminosity function leads to a change in the relative  
  number of galaxies in small/large haloes. Second, although 
  the correlation length as a function of luminosity was used 
  as input in our fitting of the conditional luminosity function, 
  there is some freedom for the `best-fit' values of the correlation 
  lengths to change in the fitting, because the errorbars
  on the observed correlation lengths are quite large.} 

\section{Populating Haloes with Galaxies}
\label{sec:method}

\subsection{Numerical Simulations}
\label{sec:simulations}

The  main goal  of  this paper  is to  use  the CLF  described in  the
previous  Section to construct  mock galaxy  redshift surveys,  and to
study a  number of statistical properties of  these distributions that
can  be  compared  with  observations  from  existing  or  forthcoming
redshift surveys.  The distribution  of dark matter haloes is obtained
from a set of large  $N$-body simulations (dark matter only).  The set
consists of a total of  six simulations with $N=512^3$ particles each,
that have been carried out on the VPP5000 Fujitsu supercomputer of the
National    Astronomical    Observatory     of    Japan    with    the
vectorized-parallel P$^3$M code (Jing  \& Suto 2002).  Each simulation
evolves the distribution  of the dark matter from  an initial redshift
of  $z=72$ down to  $z=0$ in  a $\Lambda$CDM  `concordance' cosmology.
All simulations  consider boxes with periodic  boundary conditions; in
two  cases  $L_{\rm  box}=100   h^{-1}  \Mpc$  while  the  other  four
simulations  all  have   $L_{\rm  box}=300  h^{-1}  \Mpc$.   Different
simulations  with  the  same   box  size  are  completely  independent
realizations and are  used to estimate errors due  to cosmic variance.
The  particle  masses are  $6.2  \times  10^8  \msunh$ and  $1.7\times
10^{10} \msunh$ for the small and large box simulations, respectively.
One  of the simulations  with $L_{\rm  box}=100 \mpch$  has previously
been  used by  Jing \&  Suto  (2002) to  derive a  triaxial model  for
density profiles of CDM haloes, and  we refer the reader to that paper
for complementary information about  the simulations.  In what follows
we refer to simulations with $L_{\rm box}=100 h^{-1} \Mpc$ and $L_{\rm
box}=300  h^{-1}   \Mpc$  as  $L_{100}$   and  $L_{300}$  simulations,
respectively.
\begin{figure*}
\centerline{\psfig{figure=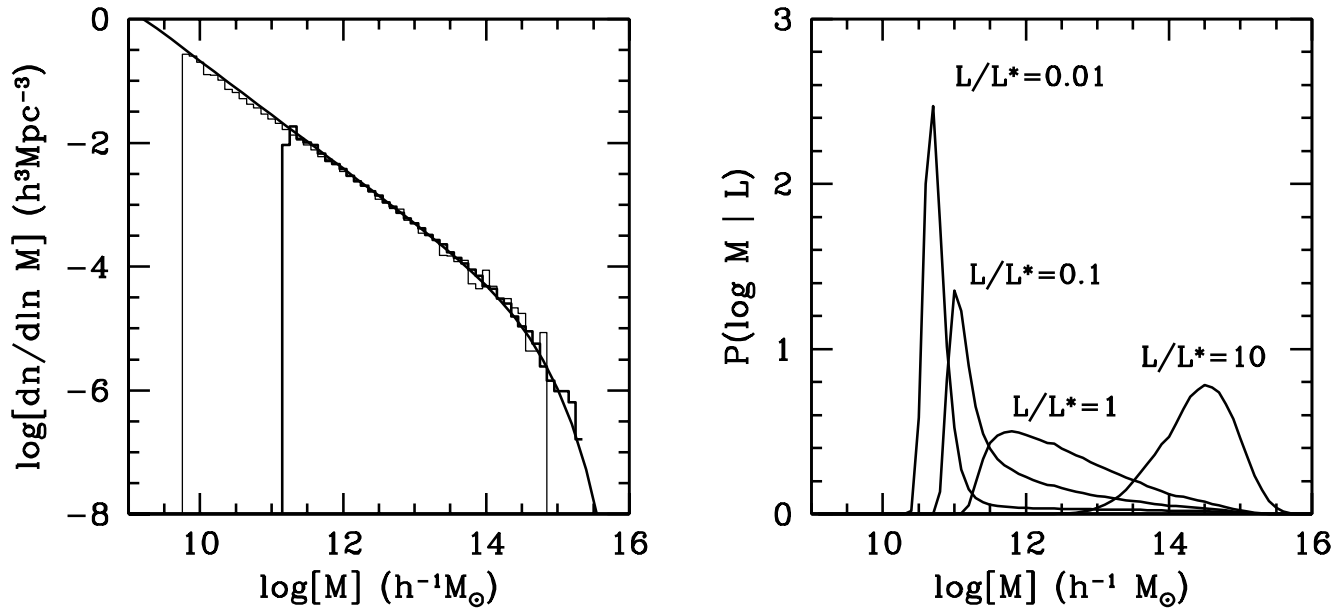,width=\hdsize}}
\caption{The  left-hand panel  plots the  halo mass  functions  of the
numerical simulations  discussed in  the text (histograms).   The mass
function with a low mass cut  at about $2 \times 10^{11} h^{-1} \Msun$
corresponds  to a  simulation with  $L_{\rm box}  = 300  h^{-1} \Mpc$,
while  the other corresponds  to a  $L_{100}$ simulation  with $L_{\rm
box} = 100  h^{-1} \Mpc$.  The solid curve is the  Sheth, Mo \& Tormen
(2001) mass function which is shown for comparison. Note the excellent
agreement, both between the two simulations and between the simulation
results and the theoretical prediction. 
The right-hand panel plots the
conditional  probability   distributions  $P(M  \vert   L)$  for  four
different  luminosities as  indicated.   $L^{*} =  1.1 \times  10^{10}
h^{-2} \Lsun$ is the characteristic luminosity of the Schechter fit to
the 2dFGRS  LF of Madgwick  \etal (2002). Combining  these conditional
probability distributions  with the halo  mass functions shown  in the
left-hand panel gives an indication of the completeness level that can
be  obtained with both  the $L_{100}$  and $L_{300}$  simulations (see
text).}
\label{fig:mf}
\end{figure*}

Dark    matter   haloes    are   identified    using    the   standard
friends-of-friends (FOF) algorithm (Davis \etal  1985) with a linking length
of $0.2$  times the  mean inter-particle separation.   Haloes obtained
with  this  linking length  have  a  mean  overdensity of  $\sim  180$
(Porciani,  Dekel  \&  Hoffman  2002),  in  good  agreement  with  the
definition  of  halo  masses  used  in our  CLF  analysis.   For  each
individual  simulation we construct  a catalogue  of haloes  with $10$
particles or more, for which  we store the mass (number of particles),
the position of the most  bound particle, and the halo's mean velocity
and velocity  dispersion. 
Note that the FOF algorithm can sometimes select poor systems
(those with small number of particles) that are spurious 
and have abnormally large velocity dispersions. We therefore 
have made a check to make sure that the particles assigned to
a system according to the FOF algorithm are gravitationally 
bound. Our test showed that this correction is important only for 
low-mass haloes, and that it has almost no effect on our results.
The  left panel of  Fig.~\ref{fig:mf} plots
the $z=0$ halo mass functions for one of the $L_{100}$ simulations and
for one of the $L_{300}$ simulations (histograms), with all 
spurious haloes erased. For comparison, we
also  plot (solid  line) the  analytical halo  mass function  given in
Sheth \& Tormen (1999)  and Sheth, Mo \& Tormen (2001)\footnote{This
same  mass  function  is  used   in  the  CLF  analysis  described  in
Section~\ref{sec:clf}.}.   The  agreement  is  remarkably  good,  both
between the two simulations and between the simulation results and the
theoretical prediction. 

Note  that our  choice for  box sizes  of $100  h^{-1} \Mpc$  and $300
h^{-1}  \Mpc$ is  a  compromise  between high  mass  resolution and  a
sufficiently  large volume  to  study the  large-scale structure.  The
impact of mass resolution is apparent from considering the conditional
probability function
\begin{equation}
\label{probM}
P(M \vert L) \, {\rm d}M = {\Phi(L \vert M) \over
\Phi(L)} \, n(M) \, {\rm d}M\,,
\end{equation}
(see Paper~I), which gives the probability that a galaxy of luminosity
$L$ resides in a halo with mass  in the range $M \pm {\rm d}M/2$.  The
right panel  of Fig.~\ref{fig:mf} plots  this probability distribution
obtained  from  the  CLF   given  in  Section~\ref{sec:clf}  for  four
different luminosities: $L = L^{*}/100$,  $L = L^{*}/10$, $L = L^{*}$,
and  $L = 10  \, L^{*}$.   Whereas $10  L^{*}$ galaxies  are typically
found  in haloes  with  $10^{13}  h^{-1} \lta  M  \lta 10^{15}  h^{-1}
\Msun$,  galaxies with  $L =  L^{*} /  100 \sim  10^{8}  h^{-2} \Lsun$
typically  reside in  haloes of  $M \simeq  5 \times  10^{10} \msunh$.
Comparing these probability distributions with the halo mass functions
in  the left panel,  we see  that the  $L_{300}$ simulations  can only
yield a {\it complete} galaxy distribution down to $L \sim 0.4 L^{*}$.
The $L_{100}$ simulation, however, resolves dark matter haloes down to
masses of  $10^{10} h^{-1}  \Msun$, which is  sufficient to  model the
galaxy population  down to  $L \sim 0.01  L^{*}$.  On the  other hand,
luminous  galaxies   may  be  under-represented  in   this  small  box
simulation, because  it contains  fewer massive haloes  than expected.
Combining these  two sets of  simulations, however, will enable  us to
study the  clustering properties  of galaxies covering  a sufficiently
large volume and a sufficiently large range of luminosities.
\begin{figure*}
\centerline{\psfig{figure=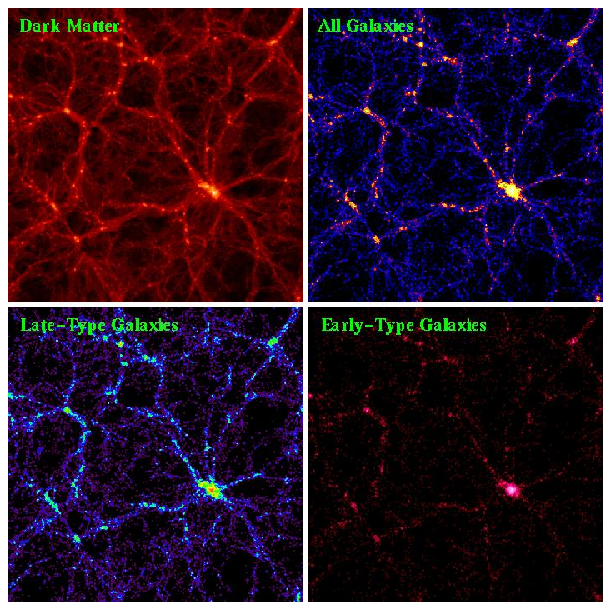,width=0.9\hdsize}}
\caption{Projected dark  matter/galaxy distributions of  a $100 \times
100 \times 10  h^{-1} \Mpc$ slice in one of  the $L_{100}$ mock galaxy
distributions.   The panels  show (clockwise  from top-left)  the dark
matter particles, all galaxies (early plus late), early-type galaxies,
and late-type  galaxies. Galaxies are weighted  by their luminosities.
Note  how the galaxies  trace the  large scale  structure of  the dark
matter, and  how early-type galaxies are more  strongly clustered than
late-type galaxies.}
\label{fig:slice1}
\end{figure*}
\begin{figure*}
\centerline{\psfig{figure=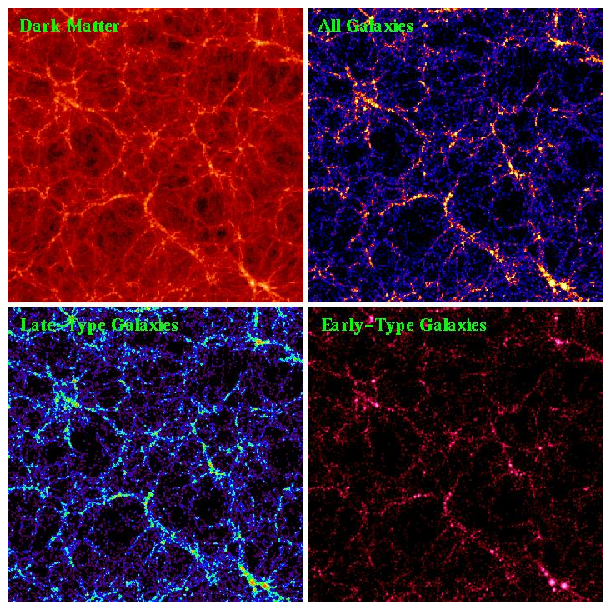,width=0.9\hdsize}}
\caption{Same  as Fig.~\ref{fig:slice1},  but  for a  $300 \times  300
\times  20 h^{-1} \Mpc$  slice taken  from one  of the  $L_{300}$ mock
galaxy distributions.}
\label{fig:slice2}
\end{figure*}

\subsection{Halo Occupation Numbers}
\label{sec:hon}

When populating haloes with galaxies  based on the CLF one first needs
to choose a  minimum luminosity.  Based on the  mass resolution of the
simulations we  adopt $L_{\rm min} = 10^{7}  h^{-2} \Lsun$ throughout.
The {\it mean} occupation number of galaxies with $L \geq L_{\rm min}$
for a halo with mass $M$ then follows from the CLF according to:
\begin{equation}
\label{meanN}
\langle N(M) \rangle = \int_{L_{\rm min}}^{\infty} \Phi(L \vert M) \,
{\rm d}L\,.
\end{equation}
In  order  to Monte-Carlo  sample  occupation  numbers for  individual
haloes one  requires the full probability distribution  $P(N \vert M)$
(with $N$ an integer) of  which $\langle N(M) \rangle$ gives the mean,
i.e.,
\begin{equation}
\label{meanNint}
\langle  N(M) \rangle = \sum_{N=0}^{\infty} N \, P(N \vert M)
\end{equation}
As a simple model we adopt
\begin{equation}
\label{pnm}
P(N \vert M) = \left\{ \begin{array}{lll}
N' + 1 - \langle N(M) \rangle & \mbox{if $N = N'$} \\
\langle N(M) \rangle - N'     & \mbox{if $N = N' + 1$} \\
0 & \mbox{otherwise}
\end{array} \right.
\end{equation}
Here $N'$ is the largest  integer smaller than $\langle N(M) \rangle$.
Thus, the  actual number of galaxies in  a halo of mass  $M$ is either
$N'$ or  $N'+1$.  This particular  model for the distribution  of halo
occupation  numbers   is  supported  by   semi-analytical  models  and
hydrodynamical simulations of  structure formation (Benson \etal 2000;
Berlind  \etal   2003)  which   indicate  that  the   halo  occupation
probability distribution is narrower  than a Poisson distribution with
the same mean.  In addition, distribution~(\ref{pnm}) is successful in
yielding  power-law  correlation  functions,  much more  so  than  for
example a Poisson distribution (Benson \etal 2000; Berlind \& Weinberg
2002).
\begin{figure*}
\centerline{\psfig{figure=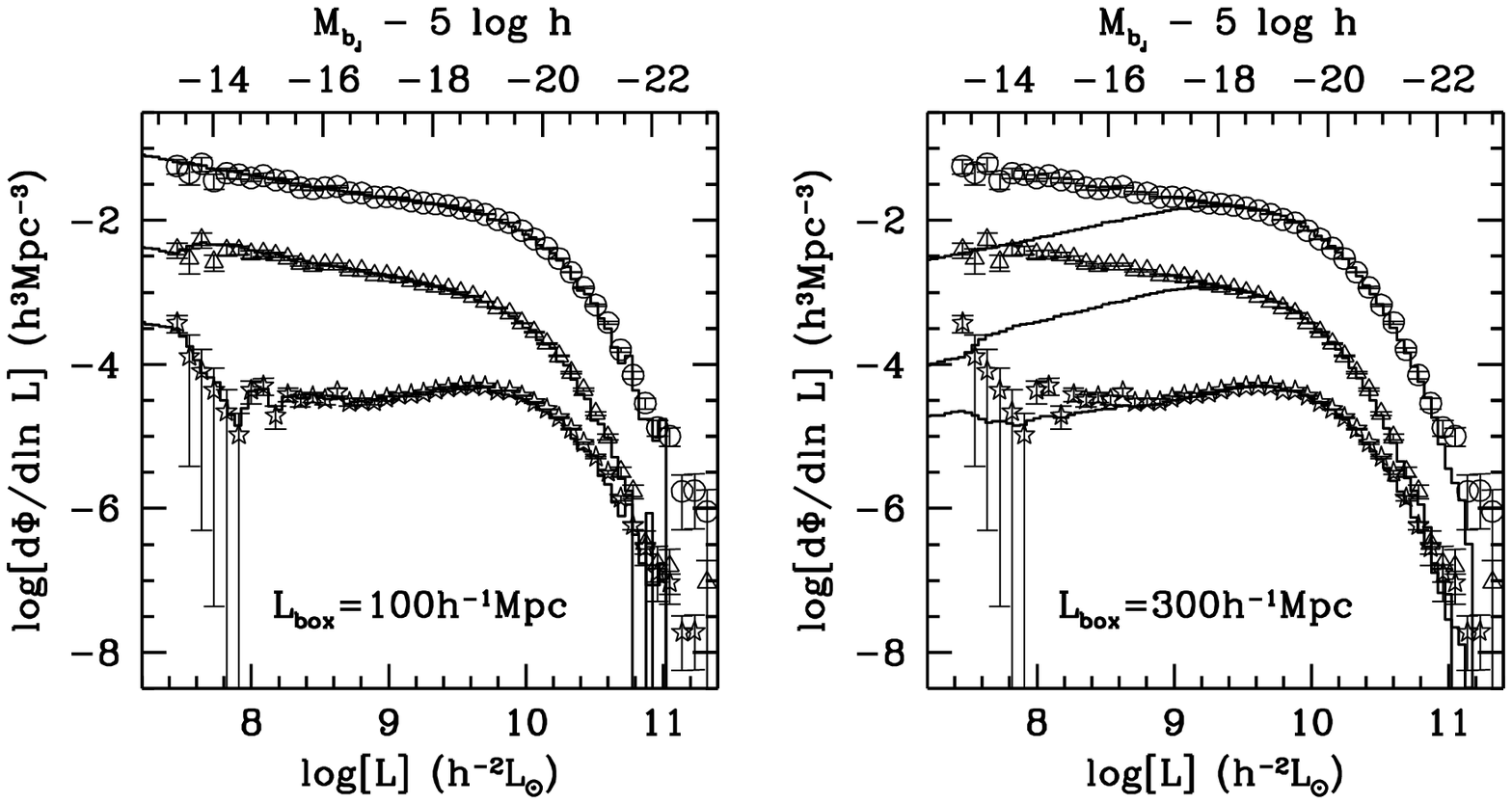,width=\hdsize}}
\caption{The  luminosity functions  of the  mock  galaxies constructed
from the $L_{100}$ (left) and $L_{300}$ (right) halo catalogues (solid
lines).  For  comparison, we  also plot the  LFs obtained  by Madgwick
\etal  (2002)  for  all  galaxies (circles),  for  late-type  galaxies
(triangles)  and for  early-type galaxies  (stars).  For  clarity, the
latter  two LFs  have  been shifted  down  by one  and  two orders  of
magnitude   in   the    $y$-direction,   respectively.    Except   for
incompleteness effects due  to the sampling of the  halo mass function
(see text  for details), the  mock galaxy distributions have  LFs that
are in excellent agreement with the data.}
\label{fig:lf}
\end{figure*}
\begin{figure}
\centerline{\psfig{figure=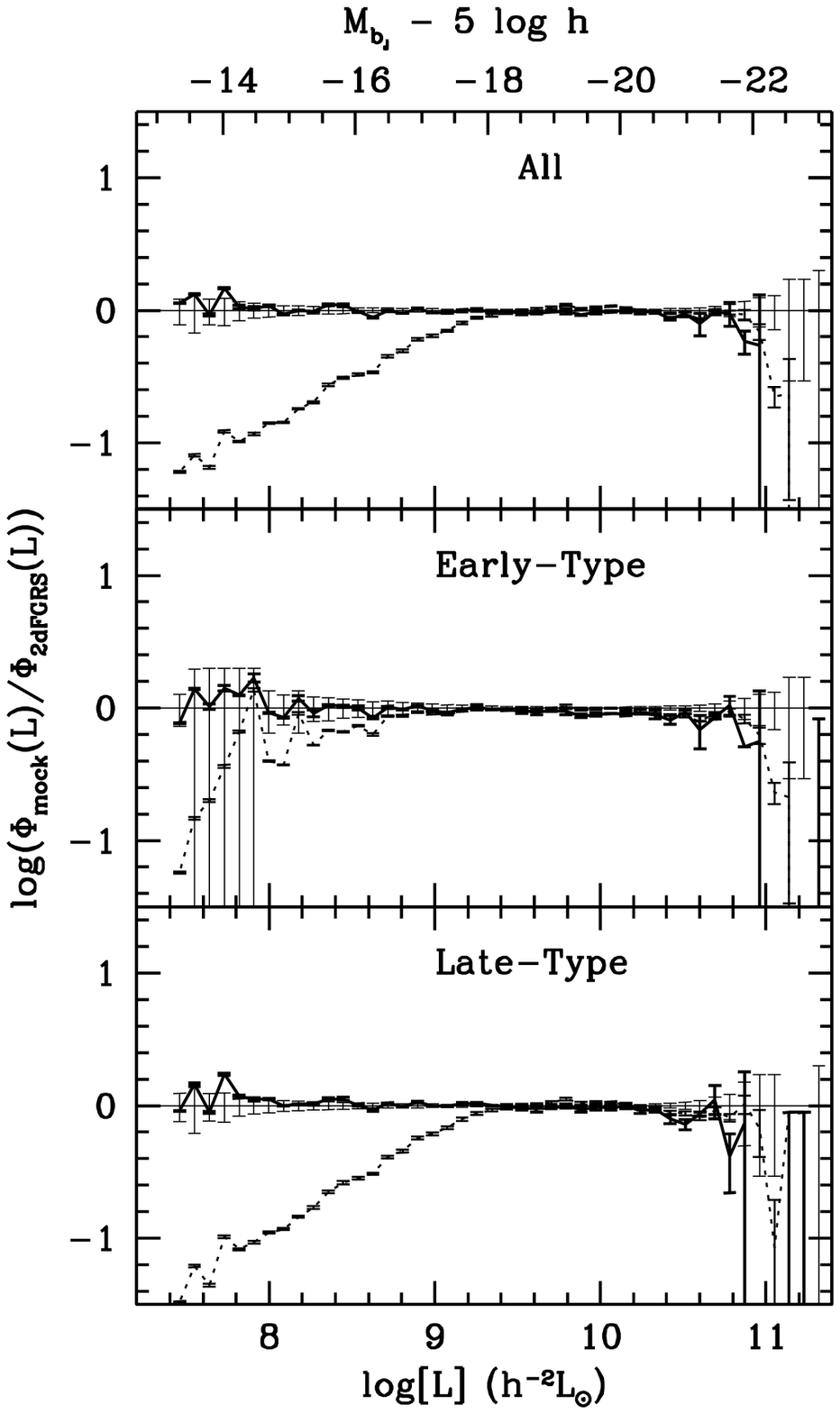,width=\hssize}}
\caption{The  ratio  of  the  luminosity function  of  mock  galaxies,
$\Phi_{\rm mock}(L)$,  to that  of the 2dFGRS,  $\Phi_{\rm 2dFGRS}(L)$
(taken  from Madgwick \etal  2002).  The  thin errorbars  indicate the
errors  on $\Phi_{\rm  2dFGRS}(L)$.   The thick  solid (dashed)  lines
correspond  to  the  LFs   obtained  from  the  $L_{100}$  ($L_{300}$)
simulations. The errorbars for the mock galaxies are obtained from the
1-$\sigma$  variance  of the  two  $L_{100}$  and  the four  $L_{300}$
simulations, respectively. See text for discussion.}
\label{fig:lfratio}
\end{figure}

\subsection{Assigning galaxies their luminosity and type}
\label{sec:luminosity}

Since the  CLF only  gives the {\it  average} number of  galaxies with
luminosities in  the range $L \pm {\rm  d}L/2$ in a halo  of mass $M$,
there are many different ways  in which one can assign luminosities to
the $N_i$  galaxies of halo $i$,  and yet be consistent  with the CLF.
The simplest approach would be to simply draw $N_i$ luminosities (with
$L >  L_{\rm min}$) randomly  from $\Phi(L \vert  M)$.  Alternatively,
one could use a more deterministic approach, and, for instance, always
demand that the $j^{\rm th}$  brightest galaxy has a luminosity in the
range $[L_{j},L_{j-1}]$.  Here  $L_j$ is defined such that  a halo has
on average $j$ galaxies with $L > L_j$, i.e.,
\begin{equation}
\label{Lj}
\int_{L_j}^\infty \Phi(L\vert M) dL = j\,.
\end{equation}
We adopt  an intermediate approach  in most of our  discussion, giving
special  treatment only  to the  one brightest  galaxy per  halo.  The
luminosity of this so-called  ``central'' galaxy, $L_c$, is drawn from
$\Phi(L\vert  M)$  with  the  restriction  $L>L_1$  and  thus  has  an
expectation value of
\begin{equation}
\label{Lcentral}
\langle L_c (M) \rangle = \int_{L_1}^{\infty}
\Phi(L \vert M) \, L \, {\rm d}L = \tilde{\Phi}^{*} \, \wLstar \,
\Gamma(\walpha+2,L_1/\wLstar),
\end{equation}
The  remaining  $N_i-1$  galaxies  are referred  to  as  ``satellite''
galaxies and are assigned luminosities in the range $L_{\rm min} < L <
L_1$, again drawn from the distribution function $\Phi(L\vert M)$.  In
Section 4.2,  we test the  effect of luminosity sampling  by comparing
the results obtained from all the three approaches.

Finally, the galaxies are assigned morphological types as follows. For
each galaxy with luminosity $L$ in a halo of mass $M$ we draw a random
number  ${\cal  R}$  in the  range  $[0,1]$.  If  ${\cal R}  <  f_{\rm
late}(L,M)$ then the galaxy is a late-type, otherwise an early-type.

\subsection{Assigning galaxies their phase-space coordinates}
\label{sec:position}

Once  the population of  galaxies has  been assigned  luminosities and
types, they need  to be assigned a position within  their halo as well
as a peculiar velocity. The central galaxy is assumed to be located at
the ``center'' of the corresponding dark halo, which we associate with
the position of the most  bound particle, and its peculiar velocity is
set equal to the mean  halo velocity (cf.  Yoshikawa, Jing \& B\"orner
2003).  For the satellite galaxies we follow two different approaches.
In  the first,  we assign  the  $N_i-1$ satellites  the positions  and
peculiar  velocities  of  $N_i  -  1$ randomly  selected  dark  matter
particles that are part of the FOF halo under consideration. This thus
corresponds to  a scenario in which satellite  galaxies are completely
unbiased with respect to the density and velocity distribution of dark
matter  particles  in  FOF  haloes.  We refer  to  satellite  galaxies
populated this way as ``FOF satellites''.

We  also   consider  a  more   analytical  model  for   the  satellite
distribution. This allows  us first of all to  assess whether a simple
analytical  description can  be found  to describe  the  population of
satellite galaxies, and secondly,  provides us with a simple framework
to  investigate the  sensitivity of  various clustering  statistics to
details regarding the density and velocity bias of satellite galaxies.
We assume  that the number density distribution  of satellite galaxies
follows a NFW density distribution (Navarro, Frenk \& White 1997):
\begin{equation}
\label{NFW}
\rho(r) = \frac{\bar{\delta}\bar{\rho}}{(r/r_{\rm s})(1+r/r_{\rm  s})^{2}},
\end{equation}
where $r_s$  is a characteristic  radius, $\bar{\rho}$ is  the average
density  of  the  Universe,  and  $\bar{\delta}$  is  a  dimensionless
amplitude which  can be expressed  in terms of the  halo concentration
parameter $c=r_{180}/r_s$ as
\begin{equation}
\label{overdensity}
\bar{\delta} = {180 \over 3} \, {c^{3} \over {\rm ln}(1+c) - c/(1+c)}.
\end{equation}
Here $r_{180}$ is  the radius inside of which the  halo has an average
overdensity   of  $180$.    Numerical  simulations   show   that  halo
concentration depends on  halo mass, and we use  the relation given by
Bullock  \etal  (2001),  converted  to  the $c$  appropriate  for  our
definition of  halo mass.  The  radial number density  distribution of
satellite galaxies  is assumed  to follow equation~(\ref{NFW})  with a
concentration  $c_g=c$, and  the  angular position  is  assumed to  be
random over  the $4\pi$ solid angle.  Peculiar  velocities are assumed
to be the sum of the peculiar  (mean) velocity of the host halo plus a
random velocity  which is assumed to be  distributed isotropically and
to follow a Gaussian, one-dimensional velocity distribution:
\begin{equation}
\label{gasv}
f(v_j)={1\over \sqrt{2\pi}\sigma_{\rm gal}}
\exp \left( -{v_j^2 \over 2\sigma_{\rm gal}^2} \right).
\end{equation}
Here  $v_j$ is the  velocity relative  to that  of the  central galaxy
along axis $j$, and $\sigma_{\rm gal}$ is the one-dimensional velocity
dispersion of  the galaxies, which  we set equal  to that of  the dark
matter particles, $\sigma_{\rm dm}$,  in the halo under consideration.
We  refer   to  satellite  galaxies   populated  this  way   as  ``NFW
satellites''.

\begin{figure*}
\centerline{\psfig{figure=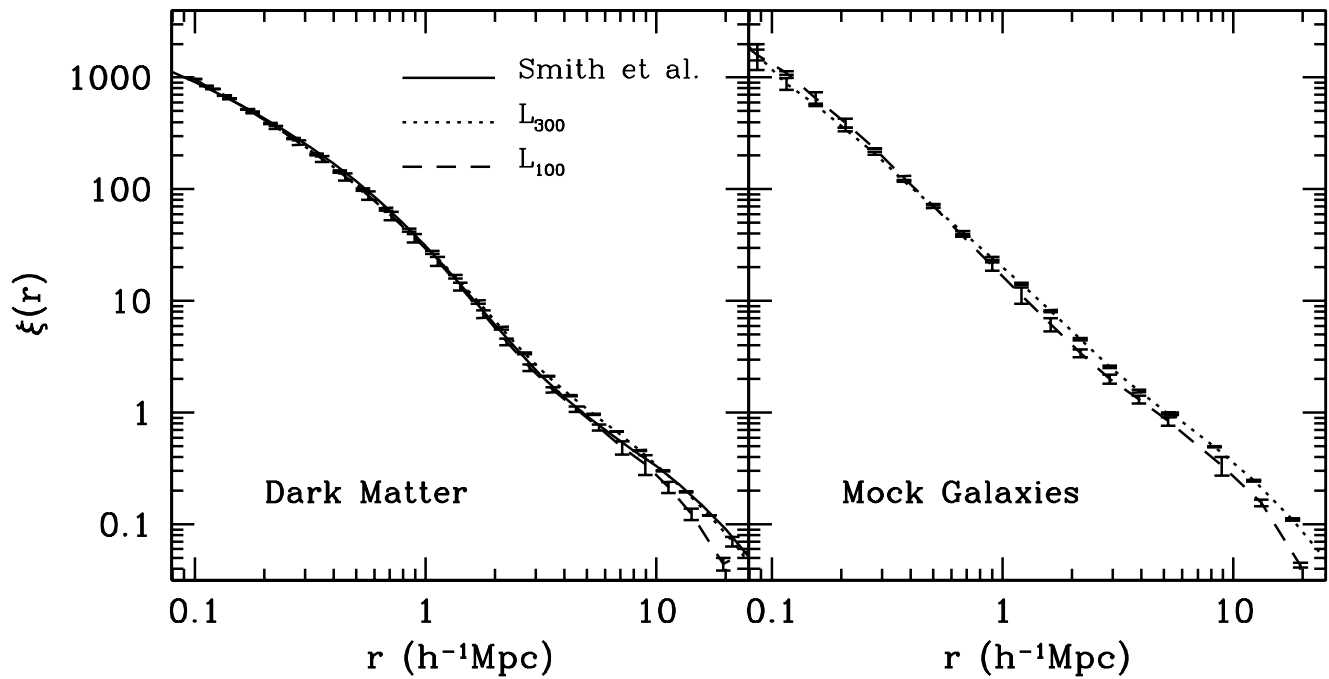,width=\hdsize}}
\caption{Two  point correlation  functions for  dark  matter particles
(left panel)  and mock galaxies  (right panel). The dotted  and dashed
lines  correspond   to  results  from  the   $L_{300}$  and  $L_{100}$
simulations,  respectively.    The  solid  line  in   the  left  panel
corresponds to  the evolved,  non-linear correlation function  for the
dark  matter  obtained  by  Smith  \etal  (2003),  and  is  shown  for
comparison.  Due  to the limited box-sizes,  the $L_{300}$ ($L_{100}$)
simulations  slightly over  (under) predict  the correlation  power on
large scales with  respect to Smith et al.'s model.   The 2PCFs in the
right  panel  are calculated  for  galaxies  with absolute  magnitudes
$M_{b_J} -  5 \log h <  -18.4$, which corresponds  to the completeness
limit of the  $L_{300}$ MGDs. Note that the box  size also affects the
2PCFs of the mock galaxies on large scales. Errorbars are the variance
among   the   two  ($L_{100}$)   and   four  ($L_{300}$)   independent
realizations.}
\label{fig:2pcfdm}
\end{figure*}

\section{Results in Real Space}
\label{sec:resreal}

Fig.~\ref{fig:slice1} and~\ref{fig:slice2} show  slices of mock galaxy
distributions   (hereafter  MGDs)   constructed  from   $L_{100}$  and
$L_{300}$ simulations,  respectively. Satellite galaxies  are assigned
positions and velocities using  the NFW scheme outlined above. Results
are shown  for all galaxies  (upper right panels), and  separately for
early types (lower  right panels) and late types  (lower left panels).
For comparison, we also show the distribution of dark matter particles
in the upper  left panels.  Note how the large  scale structure in the
dark  matter  distribution  is   delineated  by  the  distribution  of
galaxies,  and that  early-type galaxies  are more  strongly clustered
than late-type galaxies.

In this section we discuss the general, {\it real-space} properties of
these MGDs.   In Section~\ref{sec:red} below we  construct mock galaxy
redshift surveys  to investigate  the impact of  redshift distortions.
The main  goal of this section,  however, is to  investigate with what
accuracy the combination of numerical simulations and our CLF analysis
can be used to construct self-consistent mock galaxy distributions. In
particular, we want to examine to what accuracy these MGDs can recover
the input used to constrain the CLFs.  Note that this is not a trivial
question.  The  CLF modeling  is based on  the halo model,  which only
yields   an  {\it   approximate}  description   of  the   dark  matter
distribution in  the non-linear regime  (see discussions in  Cooray \&
Sheth  2002  and  Huffenberger  \&  Seljak  2003).   In  addition,  as
described in  Section~\ref{sec:method}, the  CLF alone does  not yield
sufficient  information  to  construct   MGDs,  and  we  had  to  make
additional assumptions  regarding the distribution  of galaxies within
individual haloes.  A  further goal of this section,  therefore, is to
investigate how these assumptions impact on the clustering statistics.

\subsection{The luminosity function}
\label{sec:lf}

The  CLFs used to  construct the  MGDs shown  in Fig.~\ref{fig:slice1}
and~\ref{fig:slice2}   are  constrained   by  the   2dFGRS  luminosity
functions for early- and late-type galaxies obtained by Madgwick \etal
(2002). Therefore, as  long as the halo mass  function is well sampled
by the simulations, the LFs of our MGDs should match those of Madgwick
\etal (2002). Fig.~\ref{fig:lf} shows  a comparison between the 2dFGRS
LFs  (symbols with  errorbars) and  the ones  recovered from  the MGDs
(solid  lines).   To emphasize  the  level  of  agreement between  the
recovered  LFs and  the  input LFs,  Fig.~\ref{fig:lfratio} plots  the
ratio  between the  two.   Over  a large  range  of luminosities,  the
recovered LFs  match the observational  input extremely well.   In the
$L_{300}$ simulation, however, the LFs are under-estimated for $L \lta
3 \times 10^{9} \Lsunhh$ ($M_{b_J}-5\log h \gta -18.4$).  This owes to
the absence  of haloes with  $M \lta 2  \times 10^{11}h^{-1}M_{\odot}$
(see Fig.~\ref{fig:mf}).  Note how  this discrepancy sets in at higher
$L$ for the  late-type galaxies than for the  early-types, because the
latter  are preferentially located  in more  massive haloes.   For the
early-types  the   $L_{300}$  mock  is  virtually   complete  down  to
$M_{b_J}-5\log h \simeq -17$ (see Fig.~10 of Paper II), reflecting the
fact  that only  a  very  small fraction  of  the early-type  galaxies
brighter  than  this  magnitude   reside  in  haloes  below  the  mass
resolution limit. In the $L_{100}$ simulations, on the other hand, the
LFs accurately match  the data down to the  faintest luminosities, but
here the MGD underestimates the  LFs at the bright end ($M_{b_J}-5\log
h  \lta -22$).  This  owes to  the limited  boxsize, which  causes the
number of massive haloes (the main hosts of the brightest galaxies) to
be underestimated (cf.  Fig.~\ref{fig:mf}).  Note that even the LFs of
the $L_{300}$  simulations underestimate the {\it  observed} number of
bright  galaxies.  This,  reflects a  small inaccuracy  of our  CLF to
accurately match the observed bright end of the LFs (see paper~II).
\begin{figure}
\centerline{\psfig{figure=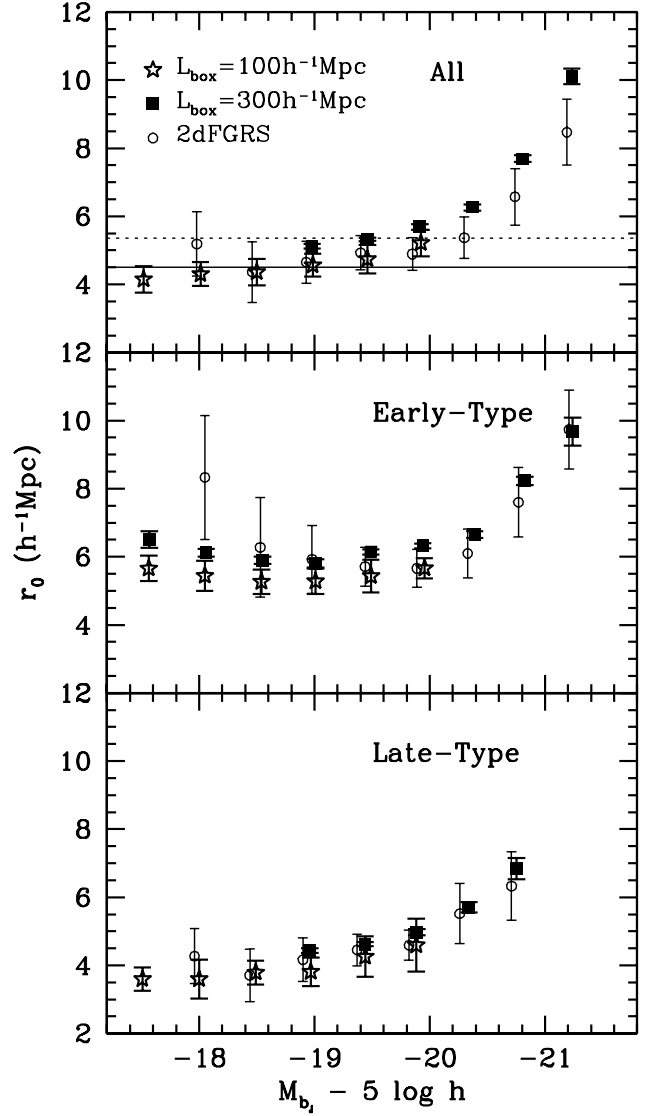,width=\hssize}}
\caption{The real  space correlation length,  $r_0$, as a  function of
galaxy  luminosity and  type.  Top  panel  shows the  results for  the
combined sample  of early- plus  late-type galaxies, while  the middle
(bottom)  panel  shows results  for  the  early  (late) type  galaxies
only. Solid  squares and stars  correspond to the  correlation lengths
obtained    from   the    $L_{300}$    and   $L_{100}$    simulations,
respectively. The errorbars correspond to the 1-$\sigma$ variance from
the two (four) independent realizations for $L_{100}$ ($L_{300}$).  We
also indicate  (open circles  with errorbars) the  correlation lengths
obtained from the 2dFGRS by  Norberg \etal (2002a). In the upper panel,
we also  plot the  correlation lengths for  dark matter  particles for
$L_{100}$  (solid  line)  and  $L_{300}$  (dotted  line)  simulations.
Although the agreement  between data and MGDs is  reasonable there are
small but significant differences.  The reason for these discrepancies
is discussed in the text.} \label{fig:r0}
\end{figure}

\subsection{The real-space correlation function}
\label{sec:2pcf}

In addition to the LFs of early- and late-type galaxies, the CLFs used
here to construct our MGDs  are also constrained by the luminosity and
type dependence of the correlation lengths as measured from the 2dFGRS
by Norberg \etal  (2002a). Here we check to  what degree this ``input''
is recovered from the MGDs.  

The left panel of Fig.~\ref{fig:2pcfdm} plots the real-space two-point
correlation  functions  (2PCFs)  for  dark  matter  particles  in  the
$L_{100}$  (dashed line) and  $L_{300}$ (dotted line) simulations.
The  solid line  corresponds to  the evolved,  non-linear  dark matter
correlation  function   of  Smith  \etal  (2003)  and   is  shown  for
comparison\footnote{In fitting  the CLF we have used  this function to
compute the  correlation length of  the dark matter  (see Paper~II).}.
As  one can  see, on  large scales  ($r \gta  6 h^{-1}{\rm  Mpc}$) the
correlation  amplitude  obtained  from  the $L_{100}$  simulations  is
systematically  lower  than  both  that obtained  from  the  $L_{300}$
simulations and that obtained from the fitting formula of Smith \etal,
suggesting that the box-size effect is non-negligible in the $L_{100}$
simulations.   Note also  that the  large scale  correlation amplitude
given  by the  $L_{300}$  simulations is  slightly  higher than  Smith
\etal's  model.  It  is  unclear if  this  discrepancy is  due to  the
inaccuracy of  the fitting formula, or  due to cosmic  variance in the
present simulations.   As we will  see below, this  discrepancy limits
the accuracy of model predictions.

The right-hand panel of  Fig.~\ref{fig:2pcfdm} plots the 2PCFs for the
{\it galaxies}  in the $L_{100}$  (dashed line) and  $L_{300}$ (dotted
line)    MGDs.  Note how the galaxies  reveal the same  trend on large
scales as the  dark matter particles, with larger  correlations in the
$L_{300}$ than in the $L_{100}$ MGD.

Fig.~\ref{fig:r0} shows  the correlation lengths $r_0$  as function of
luminosity  for  all  (upper  panel), early-type  (middle  panel)  and
late-type (lower panel) galaxies.  These have been obtained by fitting
$\xi(r)$  with   a  power   law  relation  of   the  form   $\xi(r)  =
(r/r_0)^{-\gamma}$ over  the same range  of scales as used  by Norberg
\etal (2002a).  Solid squares  and open stars correspond to correlation
lengths obtained from the  $L_{300}$ and $L_{100}$ MGDs, respectively.
Note that the errobars on  the predicted correlation lengths are based
on  the  scatter  among   independent  simulations  boxes.   They  are
significantly smaller  than the  errorbars on the  observational data,
because  the model  predictions  are based  on real-space  correlation
functions,  while the  observational  results are  based on  projected
correlation functions in redshift  space.  The agreement with the data
(open circles)  is reasonable,  even though several  systematic trends
are apparent.   In particular,  the correlation lengths  obtained from
the  $L_{300}$ simulation  are slightly  higher than  the observations
while  the  opposite  applies  to  the  $L_{100}$  simulation.   These
discrepancies  are due  to two  effects.  First  of all,  as  shown in
Fig.~\ref{fig:2pcfdm} the dark matter on large scales is more strongly
clustered  in   the  $L_{300}$  simulations  than   in  the  $L_{100}$
simulations. That this can account for most of the differences between
the   scale-lengths  obtained   from  the   $L_{300}$   and  $L_{100}$
simulations, is illustrated by  the dotted and solid horizontal lines,
which indicate the correlation lengths of the dark matter particles in
the $L_{300}$  and $L_{100}$ simulations,  respectively. Secondly, the
measured correlation lengths correspond to a non-zero, median redshift
which is  larger for the  more luminous galaxies.  In  determining the
best-fit parameters  for the  CLF this redshift  effect is  taken into
account (see  Papers~I and II).   However, in the construction  of our
MGDs, we only use the  dark matter distribution at $z=0$. As discussed
in Paper  I, this  can over-estimate the  correlation length  by about
$10\%$.  Given  these sources  of  systematic  errors,  one should  be
careful not to over-interpret  any discrepancy between the correlation
lengths  in the  mock survey  and  those obtained  from real  redshift
distributions.
\begin{figure}
\centerline{\psfig{figure=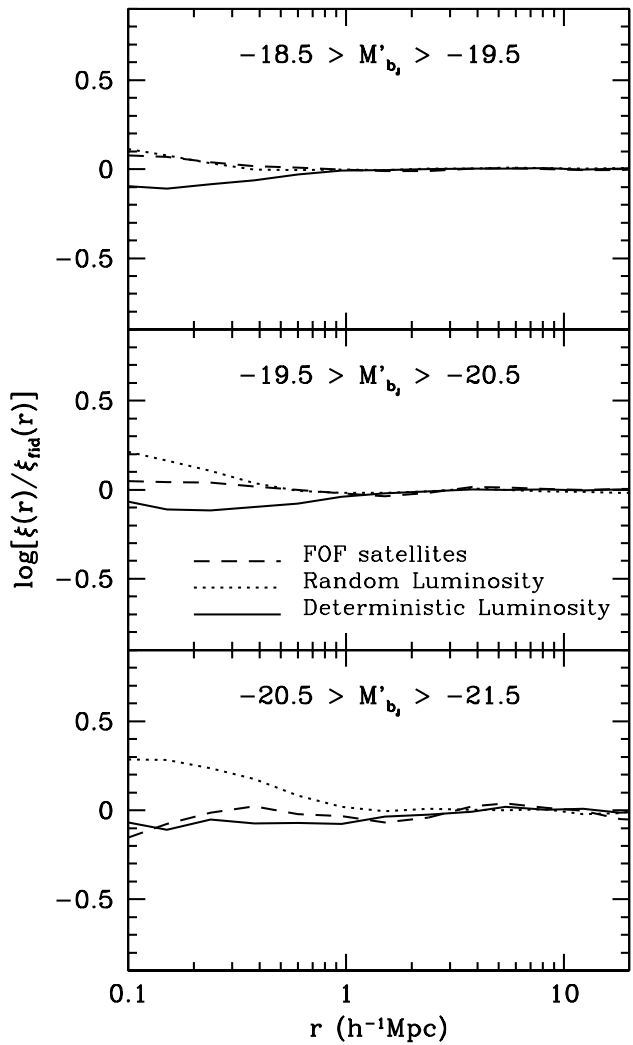,width=\hssize}}
\caption{The ratio of the 2PCF $\xi(r)$ in three MGDs compared to that
of our fiducial  MGD. The only difference among  these various MGDs is
the way that we assign luminosities and phase-space coordinates to the
galaxies. Solid (dotted)  lines correspond to a MGD in  which we use a
deterministic  (random) method to  assign galaxies  their luminosities
(see  Section~\ref{sec:luminosity}  for   definitions).   In  the  MGD
corresponding  to the dashed  line we  use the  intermediate, fiducial
method to assign luminosities, but here we use `FOF satellites' rather
than    `NFW   satellites'    (see    Section~\ref{sec:position}   for
definitions).   Results  are shown  for  galaxies  in three  different
magnitude  bins (as indicated)  in one  of the  $L_{300}$ simulations.
However,  results   for  the  $L_{100}$   simulations  look  virtually
identical.}
\label{fig:cgrg}
\end{figure}

In order to investigate the sensitivity of the 2PCF in the MGDs to the
way we assign luminosities and phase-space coordinates to the galaxies
within  the dark matter  haloes, we  construct MGDs  using one  of the
$L_{300}$  simulations  with   different  models  for  the  luminosity
assignment  and  spatial  distribution  of satellite  galaxies  within
haloes.  We have confirmed that using one of the $L_{100}$ simulations
instead  yields the same  results.  We  first test  the impact  of the
luminosity assignment.   Here, instead of  the fiducial model  for the
luminosity   assignment  (the   intermediate  approach   discussed  in
Section~\ref{sec:luminosity}),  we  use  both  the  deterministic  and
random assignments (see Section~ \ref{sec:luminosity} for definitions)
to  construct the  MGDs. In  Fig.~\ref{fig:cgrg} we  shown  the ratios
between the  correlation functions obtained from these  MGDs and those
obtained from the fiducial MGD. For bright galaxies, the deterministic
model  gives  the  lowest  amplitudes  on  small  scales  ($r  \lta  1
h^{-1}\Mpc$),   while    the   random   model    gives   the   highest
amplitudes. This is expected. The  mean number of bright galaxies in a
typical halo is not much larger than  1 and so not many close pairs of
bright  galaxies are expected  in the  deterministic model.  More such
pairs are expected in the  random model because more than one galaxies
in a  typical halo can  be assigned a  large luminosity due  to random
fluctuations. The dashed lines  in Fig.~\ref{fig:cgrg} correspond to a
MGD  with   FOF  satellites  (see   Section~\ref{sec:position}).   The
agreement of the 2PCFs between  this MGD with `FOF satellites' and our
fiducial  MGD  indicate  that  the  spherical  NFW  model  is  a  good
approximation  of  the average  density  distribution  of dark  matter
haloes.  We have also tested  the impact of changing the concentration
of galaxies, $c_g$; increasing  (decreasing) $c_g$ with respect to the
dark matter  halo concentration, $c$, increases  (decreases) the 2PCFs
on small scales ($r \lta  1 h^{-1} \Mpc$). However, even when changing
the ratio $c_g/c$ by a factor  of two, the amplitude of this change is
smaller than  the differences  resulting from changing  the luminosity
assignment.
 
All in all, changes in  the way we assign luminosities and phase-space
coordinates to the galaxies only have  a mild impact on the 2PCFs, and
only at small scales $\lta 1  h^{-1} \Mpc$.  This is in good agreement
with Berlind \& Weinberg (2002)  who have shown that these effects are
much smaller than changes in  the second moment of the halo occupation
distributions.   For example,  assuming a  Poissonian $P(N  \vert M)$,
rather than equation~(\ref{pnm}) has a much larger impact on the 2PCFs
than  any  of   the  changes  investigated  above.   As   we  show  in
Section~\ref{sec:red}   below,   with    the   $P(N   \vert   M)$   of
equation~(\ref{pnm})  we  obtain  correlation  functions that  are  in
better  agreement with observations,  providing empirical  support for
this particular occupation number distribution.

It is  interesting to note that  although small changes in  the way we
assign  luminosities and  phase-space  coordinate do  not  have a  big
impact on  the statistical measurements we are  considering here, such
changes  can lead  to quite  different results  for  other statistical
measures. As shown  in van den Bosch \etal  (2004), various statistics
of  satellite   galaxies  around  bright  galaxies  can   be  used  to
distinguish models that make  similar predictions about the clustering
on large scales.

\subsection{Pairwise velocities}
\label{sec:pvd}

The peculiar  velocities of galaxies  are determined by the  action of
the gravitational  field, and  so are directly  related to  the matter
distribution  in  the Universe.   Observationally,  the properties  of
galaxy  peculiar  velocities  are  inferred from  distortions  in  the
correlation    function.      We    defer    this     discussion    to
Section~\ref{sec:red}.  Here we derive statistical quantities directly
from the simulated peculiar velocities of galaxies.

We define the pairwise peculiar velocity of a galaxy pair as
\begin{equation}
\label{ppvdef}
v_{12}(r) \equiv [{\bf v}({\bf x}+{\bf r}) - {\bf v}({\bf x})]
\cdot \hat{\bf r},
\end{equation}
with ${\bf  v}({\bf x})$  the peculiar velocity  of a galaxy  at ${\bf
x}$.  The  mean pairwise peculiar  velocity and the  pairwise peculiar
velocity dispersion (PVD) are
\begin{equation}
\label{pvddef}
\langle v_{12} (r)\rangle
~~~~\mbox{and}~~~~
\sigma_{12}(r)\equiv \langle [v_{12}(r)-\langle
v_{12}(r)\rangle]^2\rangle^{1/2} \,,
\end{equation}
where  $\langle \cdot\cdot\cdot\rangle$  denotes an  average  over all
pairs of separation $r$.

\begin{figure*}
\centerline{\psfig{figure=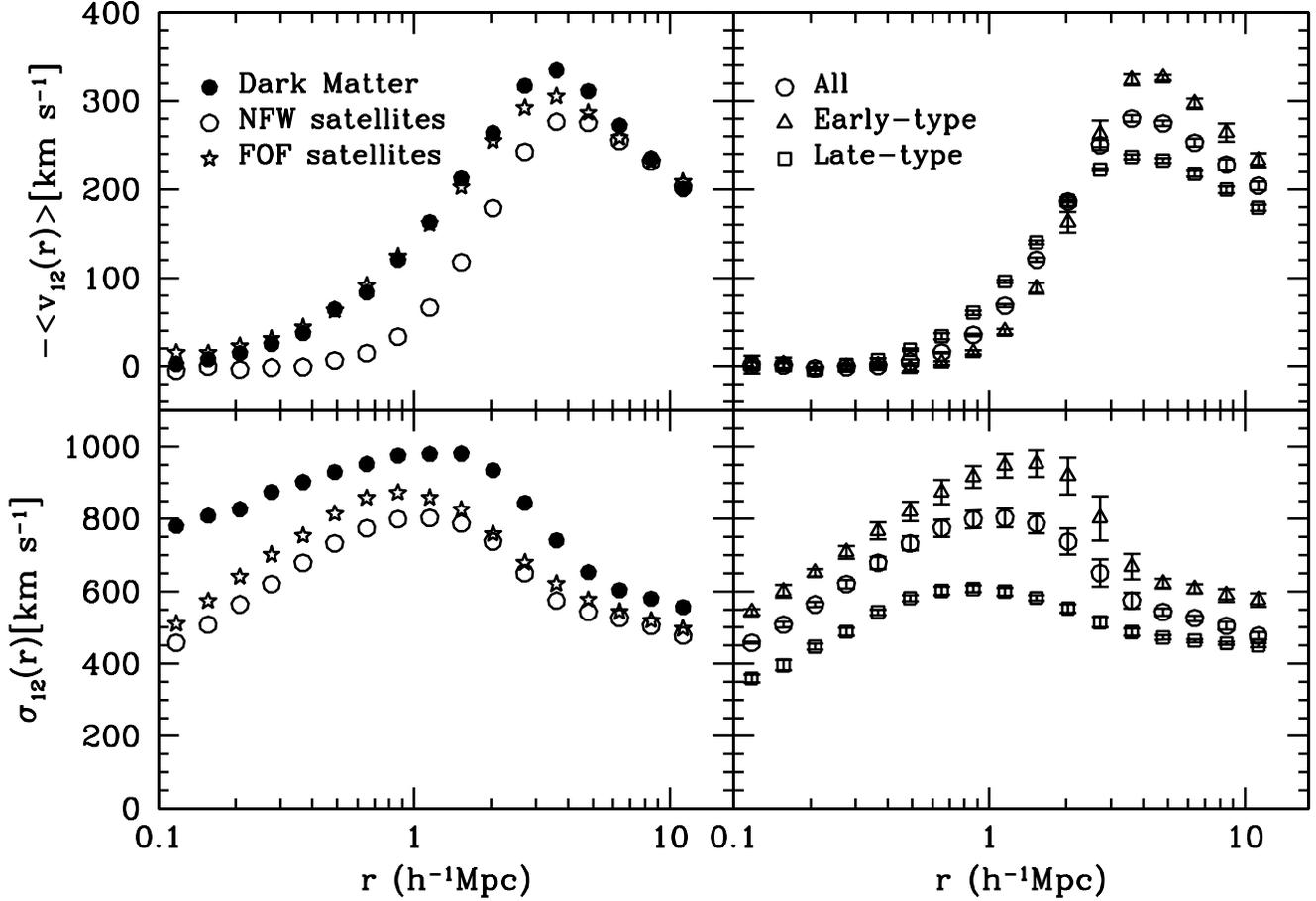,width=\hdsize}}
\caption{The  mean  pairwise velocities  (upper  panels) and  pairwise
velocity    dispersions   (lower    panels)    estimated   from    the
three-dimensional  (real-space) velocities  of the  mock  galaxies and
dark  matter  particles.   All  results correspond  to  the  $L_{300}$
simulations  only.   Left-hand panels  compare  dark matter  particles
(solid  circles)  with  galaxies  either  with  NFW  satellites  (open
circles)  or  with FOF  satellites  (open  stars).  Right-hand  panels
display  the galaxy-type dependence  for a  model with  NFW satellites
(errorbars  indicate  the   rms-scatter  among  the  four  independent
$L_{300}$ simulations).  See text for detailed discussion.}
\label{fig:pvd}
\end{figure*}
\begin{figure*}
\centerline{\psfig{figure=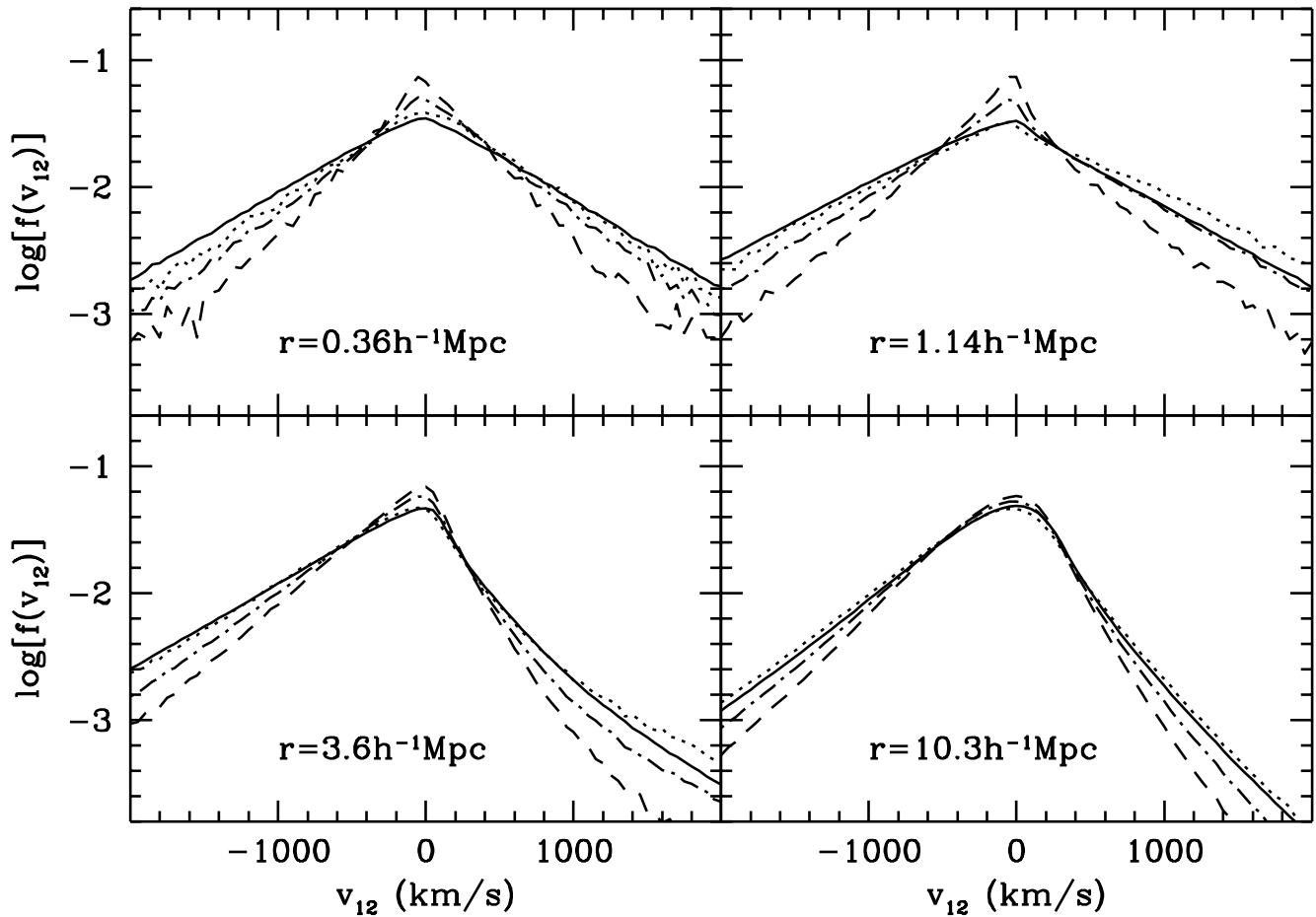,width=\hdsize}}
\caption{Distribution  of pairwise  velocities, $f(v_{12})$,  for dark
matter  particles  (solid  curves),  and  for  mock  galaxies  in  the
$L_{300}$ simulation.   Results are shown for four  separations $r$ as
indicated,  and  for   all  galaxies  (dot-dashed  lines),  early-type
galaxies  (dotted lines)  and late-type  galaxies (dashed  lines).  On
small scales ($r < 1 h^{-1} \Mpc$) the pairwise velocity distributions
are  symmetric and  reveal  an obvious  exponential  form.  On  larger
scales, however, $f(v_{12})$ reveals  clear asymmetries: for $v_{12} <
0$ the distribution is still close to an exponential, while for $v_{12}
> 0$ the distribution more resembles a normal distribution.}
\label{fig:fv}
\end{figure*}
In order to  gain insight, we compute $\langle  v_{12} (r)\rangle$ and
$\sigma_{12}(r)$ from  the $L_{300}$ simulations for  both dark matter
particles and for  galaxies with $M_{b_J} - 5 {\rm  log} h \leq -18.4$
(which corresponds to the completeness limit of these simulations, see
Fig.~\ref{fig:lf}).

Results are shown in Fig.~\ref{fig:pvd}. The upper left panel compares
the  mean pairwise peculiar  velocities of  the dark  matter particles
(solid circles)  with those of  two realizations of the  galaxies: one
with  `NFW  satellites'  (open   circles)  and  the  other  with  `FOF
satellites'  (stars).  At sufficiently  small separations,  one probes
the  virialized regions  of dark  matter  haloes, and  one thus  finds
$\langle v_{12}  \rangle =  0$. At larger  separations, one  starts to
probe  the  infall  regions  around the  virialized  haloes,  yielding
negative  values   for  $\langle  v_{12}(r)   \rangle$.   Finally,  at
sufficiently large separations  $\langle v_{12}(r) \rangle \rightarrow
0$ due to the large scale homogeneity and isotropy of the Universe.

Both the dark  matter particles and the galaxies  from our MGDs indeed
reveal such  a behavior, with  $\langle v_{12}(r) \rangle$  peaking at
$\sim 3 h^{-1} \Mpc$.  However,  there is a markedly strong difference
between the  $\langle v_{12}(r) \rangle$  of galaxies in the  MGD with
NFW satellites  and that of the  dark matter. In  this particular MGD,
the galaxies  experience significantly smaller  infall velocities than
the  dark  matter particles.  However,  this  difference between  dark
matter  and   galaxies  is   almost  absent  in   the  MGD   with  FOF
satellites. This is due to the fact that in the NFW model, we populate
satellites with  {\it isotropic} velocity dispersions  within a sphere
of radius $r_{180}$.  We are  thus assuming that the entire region out
to $r_{180}$  is virialized in that  there is no  net infall. However,
simple collapse models predict that for our concordance cosmology only
the  region out to  $r_{340}$ (i.e.,  the radius  inside of  which the
average overdensity is 340) is virialized (Bryan \& Norman 1998).  The
difference  between the MGDs  with NFW  satellites and  FOF satellites
indicates that  the regions between $r_{340}$ and  $r_{180}$ are still
infalling, resulting in non-zero $\langle v_{12} \rangle$.

In the  lower-left panel,  we compare the  PVDs for galaxies  and dark
matter particles. Here the MGDs with FOF satellites and NFW satellites
are  fairly  similar,  and  significantly  lower  than  for  the  dark
matter. This can  be understood as follows. At  small separations, the
PVD is a  {\it pair} weighted measure for the  potential well in which
dark matter particles (galaxies) reside.  For the galaxies in our MGDs
the halo occupation number per  unit mass, $N/M$, {\it decreases} with
the mass of dark matter haloes (see Paper~II).  Therefore, the massive
haloes (with  larger velocity dispersions)  contribute relatively less
to  the  PVDs  of  galaxies.   Although  the  difference  between  the
$\sigma_{12}(r)$ of  the MGDs with  FOF and NFW satellites  shows that
the  PVDs have  some dependence  on the  details regarding  the infall
regions around virialized haloes, these effects are typically small.

The upper-right and lower-right  panels of Fig.~\ref{fig:pvd} show how
$\langle  v_{12}(r)  \rangle$ and  $\sigma_{12}(r)$  depend on  galaxy
type. Results are shown for the MGD based on NFW satellites.  The mean
velocities for early-type galaxies are larger than those for late-type
galaxies on large  scales, but smaller on small  scales.  In addition,
the  PVD of  early-type  galaxies  is higher  than  that of  late-type
galaxies on all scales. All these differences are easily understood as
a reflection  of the fact that early-type  galaxies are preferentially
located in the larger, more  massive haloes which have larger velocity
dispersions and larger infall velocities.

Fig.~\ref{fig:fv} shows  the pairwise velocity  distributions for four
different separations  $r$, within  a logarithmic interval  of $\Delta
{\rm log} r  = 0.125$.  On small scales, the  distribution is well fit
by an exponential  for both dark matter particles  and galaxies.  This
validates  the   assumption  made  in  earlier   analyses  about  this
distribution (Davis \& Peebles 1983; Mo, Jing \& B\"orner 1993; Fisher
\etal 1994;  Marzke \etal 1995).   It is also consistent  with earlier
results  obtained from  theoretical models  and  numerical simulations
based on dark  matter particles (Diaferio \& Geller  1996; Sheth 1996;
Mo, Jing  \& B\"orner  1997; Seto \&  Yokoyama 1998;  Efstathiou \etal
1988; Magira, Jing \&  Suto 2000).  For larger separations $f(v_{12})$
is skewed  towards negative values of $v_{12}$,  because galaxies tend
to approach each other due  to gravitational infall. Clearly, a single
exponential function is no longer a good approximation to the pairwise
peculiar  velocity  distribution at  large  separations. Although  for
$v_{12}<0$ (infall)  the exponential remains  remarkably accurate, for
$v_{12}  >  0$  the  pairwise  velocity distribution  reveals  a  more
Gaussian  behavior.   This may  have  important  implications for  the
derivation of PVDs (especially  at large separations), which typically
is based  on the assumption  of a purely exponential  $f(v_{12})$.  We
shall return to this issue in more detail in Section~\ref{sec:pvd2}.
\begin{figure}
\centerline{\psfig{figure=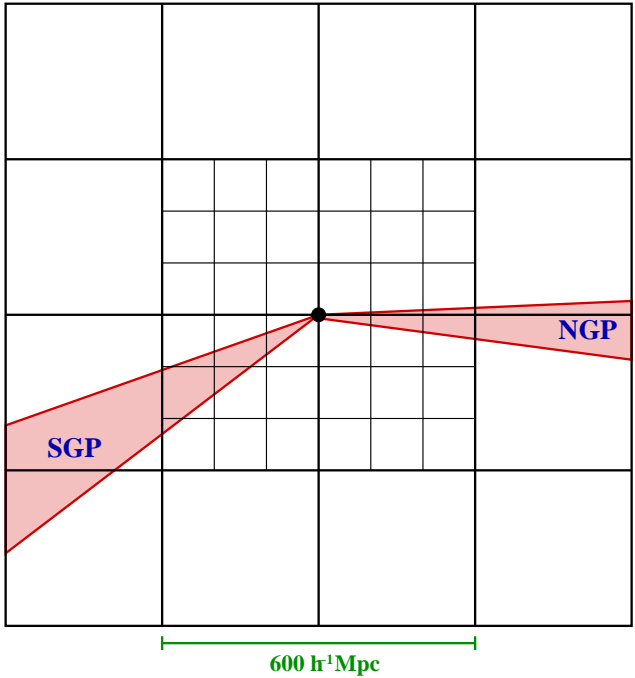,width=\hssize}}
\caption{The  stacking   geometry  of  the   $L_{100}$  and  $L_{300}$
simulation  boxes  used to  construct  the  MSB  mock galaxy redshift 
surveys. The virtual observer is located at the center of the 
stack, indicated by a thick solid dot. Note that for MGRSs
in the MB set,  the stack of $6 \times 6 \times  6$ $L_{100}$ boxes is
replaced by a stack of $2 \times 2 \times  2$ $L_{300}$ boxes.}
\label{fig:box}
\end{figure}

\section{Results in Redshift Space}
\label{sec:red}

The  statistical  quantities of  galaxy  clustering  discussed in  the
previous section are  based on real distances between  galaxies in our
MGDs. However,  because of the  peculiar velocities of  galaxies, such
quantities cannot be obtained  directly from a galaxy redshift survey.
On small scales  the virialized motion of galaxies  within dark matter
haloes cause  a reduction  of the correlation  power, while  on larger
scales  the correlations  are  boosted  due to  the  infall motion  of
galaxies towards overdensity regions (Kaiser 1987; Hamilton 1992).  As
discussed  in  the  introduction,  these  distortions  contain  useful
information  about  the  Universal  density  parameter,  the  bias  of
galaxies  on large  (linear) scales,  and the  pairwise  velocities of
galaxies.

In this  section, we use the  MGDs presented above  to construct large
mock galaxy redshift surveys (hereafter MGRSs).  The main goals are to
compare  various clustering  statistics from  these mock  surveys with
observational data from the 2dFGRS, and to investigate how the details
about the  CLF and  the distribution of  galaxies {\it  within} haloes
impact on these statistics.  For  the model-data comparison we use the
large scale structure analysis of Hawkins \etal (2003; hereafter H03),
which is based on a subsample of the 2dFGRS consisting of all galaxies
located in the North Galactic Pole (NGP) and South Galactic Pole (SGP)
survey  strips with  redshift $0.01  \leq  z \leq  0.20$ and  apparent
magnitude  $b_J  < 19.3$.  This  sample  consists  of $\sim  166,000$
galaxies covering an area on the sky of $\sim 1090 \, {\rm deg}^2$.

In  order  to  carry  out   a  proper  comparison  between  model  and
observation, we aim  to construct MGRSs that have  the same selections
as the 2dFGRS.  First of all, the survey depth of  $z_{\rm max} = 0.2$
implies  that  we  need  to  cover  a volume  with  a  depth  of  $600
h^{-1}\Mpc$,  i.e., twice that  of our  big $L_{300}$  simulations. In
principle,  we could stack  $4 \times  4\times 4$  identical $L_{300}$
boxes (which  have periodic boundary  conditions), so that a  depth of
$600 h^{-1}\Mpc$  can be  achieved in all  directions for  an observer
located at  the center  of the stack.   However, there is  one problem
with   this   set-up;  as   we   have   shown  in   Figs.~\ref{fig:mf}
and~\ref{fig:lf} the $L_{300}$ MGD is only complete down to $M_{b_J} -
5 {\rm log} h \simeq -18.4$.  Taking account of the apparent magnitude
limit of the  survey, this implies that our  MGRSs would be incomplete
out to  a distance of  $\sim 350 h^{-1}  \Mpc$.  We can  overcome this
problem by using the  higher resolution $L_{100}$ simulation, which is
complete down to  $M_{b_J} - 5 {\rm log} h  \simeq -14$.  We therefore
replace the central  $2\times 2\times 2$ $L_{300}$ boxes  with a stack
of $6  \times 6 \times  6$ $L_{100}$ boxes.  The final lay-out  of our
virtual universe is illustrated in Figure~\ref{fig:box}. Unless stated
otherwise, satellite galaxies are assigned to dark matter haloes based
on our standard NFW method described in Section~\ref{sec:position}.
 
Observational selection  effects, which are modelled  according to the
final  public data  release  of  the 2dFGRS  (see  also Norberg  \etal
2002b), are taken into account using the following steps:

\begin{enumerate}
  
\item We place a virtual observer  at the center of the stack of boxes
  (the    solid    dot     in    Figure~\ref{fig:box}),    define    a
  $(\alpha,\delta)$-coordinate frame, and remove all galaxies that are
  not located  in the areas equivalent  to the NGP and  SGP regions of
  the 2dFGRS.
  
\item Next, for  each galaxy we compute the redshift  as `seen' by the
  virtual observer.  We  take the observational velocity uncertainties
  into  account by  adding a  random  velocity drawn  from a  Gaussian
  distribution with dispersion $85\kms$ (Colless \etal 2001).
 
\item We  compute the apparent  magnitude of each galaxy  according to
  its  luminosity and  distance.  Since  galaxies in  the  2dFGRS were
  pruned  by  apparent  magnitude  {\it  before}  a  k-correction  was
  applied,  we   proceed  as  follows:  We  first   apply  a  negative
  k-correction,    then    select    galaxies   according    to    the
  position-dependent  magnitude  limit  (obtained using  the  apparant
  magnitude  limit masks  provided by  the 2dFGRS  team),  and finally
  k-correct  the  magnitudes  back  to their  rest-frame  $b_J$-band.  
  Throughout we use the type-dependent k-corrections given in Madgwick
  \etal (2002).
  
\item To  mimic the position- and  magnitude-dependent completeness of
  the 2dFGRS,  we randomly sample  each galaxy using  the completeness
  masks provided by the 2dFGRS team.  The incompleteness of the 2dFGRS
  {\it parent}  sample is taking  into account by  randomly discarding
  $9\%$ of all mock galaxies (Norberg \etal 2002b).
  
\item Finally,  we mimic the  actual selection criteria of  the 2dFGRS
  sample used in H03 by  restricting the sample to galaxies within the
  redshift range  $0.01 \leq z  \leq 0.20$ and with  completeness $\ge
  0.7$.

\end{enumerate}

Each MGRS  thus constructed  contains, on average,  $169000$ galaxies,
with a dispersion of $\sim 5000$ due to cosmic variance. The number of
galaxies in  our mock catalogues are consistent  with the observations
at the $1 \sigma$ level. Note that the correlation functions presented
by H03  have been  corrected for the  observational bias due  to fiber
collisions, and we therefore do not mimic these effects in our MGRSs.
\begin{figure*}
\centerline{\psfig{figure=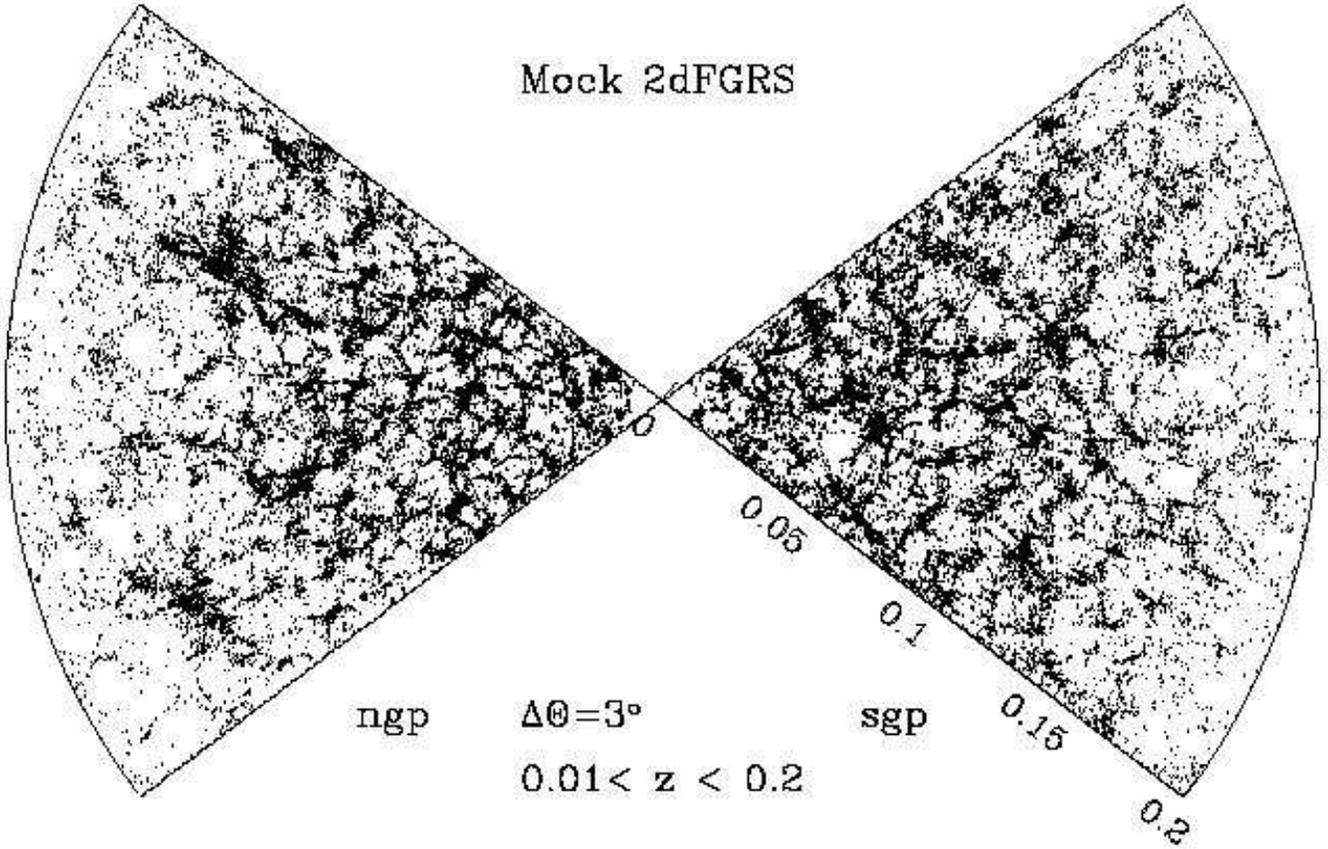,width=\hdsize}}
\caption{The distribution of a sub-set of galaxies in one of the MSB
  mock samples.  For clarity,  we plot galaxies  only in  two 3-degree
  slices, one in the `North  Galactic Pole' region (NGP) and the other
  in  the `South  Galactic  Pole' region  (SGP).   Only galaxies  with
  redshifts in the range $0.01 <z < 0.2$ are plotted.  }
\label{fig:wedge}
\end{figure*}

Since we have 2 $L_{100}$  simulations and 4 $L_{300}$ simulations, we
construct $2  \times 4=8$ mock catalogues  with different combinations
of small- and big-box simulations.   In what follows, we refer to this
set of mock  catalogues as MSBs (for Mock  Small/Big).  As an example,
Fig.~\ref{fig:wedge} shows  the distribution of a  sub-set of galaxies
in one  of these mock catalogues.   Although each of  our MSB catalogs
covers  an  extremely  large  volume,  and should  thus  not  be  very
sensitive to cosmic variance, it is constructed using simulations with
box sizes of $100$ and $300  h^{-1} \Mpc$ only.  If, for instance, the
$L_{100}$ simulation contains a big cluster, the $6 \times 6 \times 6$
reproduction of this box in our MGRSs might introduce some unrealistic
features.   Furthermore,   as  shown  in   Section~\ref{sec:2pcf}  the
$L_{100}$ box underestimates the amount of clustering on large scales.
Therefore, this set of MGRSs, which replicate this box 27 times, might
underestimate the  clustering on  large scales as  well.  In  order to
test the sensitivity  of our results to these  potential problems, and
to have a  better handle on the impact of cosmic  variance in our mock
surveys, we  construct four alternative  MGRSs. Each consists of  a $4
\times  4\times 4$  stack of  one  of the  four $L_{300}$  simulations
(i.e., we replace  the $6 \times 6 \times 6$  stack of $L_{100}$ boxes
by a $2 \times 2 \times 2$ stack of $L_{300}$ boxes).  In what follows
we refer to this set of  mock catalogues as MBs (for Mock Big).  These
MGRSs,  although incomplete for  $M_{b_J} -  5 {\rm  log} h  > -18.4$,
should not suffer  from the lack of clustering power  on large scales. 
The MSB set,  on the other hand, does  not suffer from incompleteness,
but instead lacks some large scale  power.  As we will see below, both
the MSB and MB mocks  give similar results on large scales, suggesting
that the box-size effect does  not have a significant influence on our
results.

\subsection{Two-Point Correlation Functions}
\label{sec:red2pcf}

From our MGRSs we compute $\xi(r_p,\pi)$ using the estimator (Hamilton
1993)
\begin{equation}
\label{tpcfest}
\xi(r_p,\pi) = {\langle RR \rangle \, \langle DD \rangle \over
\langle DR \rangle^2} - 1
\end{equation}
with  $\langle DD  \rangle$,  $\langle RR  \rangle$,  and $\langle  DR
\rangle$ the number of galaxy-galaxy, random-random, and galaxy-random
pairs with separation  $(r_p,\pi)$. Here $r_p$ and $\pi$  are the pair
separations   perpendicular  and   parallel   to  the   line-of-sight,
respectively.  Explicitly, for  a pair  $({\bf s_1},{\bf  s_2})$, with
${\bf s_i} = c z_i {\bf \hat{r}_i}/H_0$, we define
\begin{equation}
\label{rppi}
\pi = {{\bf s} \cdot {\bf l} \over \vert {\bf l} \vert} \; ,
\;\;\;\;\;\;\;\;\;\;\;\;\;\;
r_p = \sqrt{{\bf s} \cdot {\bf s} - \pi^2}
\end{equation}
Here ${\bf l}={1\over 2}({\bf s_1} +  {\bf s_2})$ is the line of sight
intersecting the pair,  and ${\bf s} = {\bf s_1}  - {\bf s_2}$. Random
samples are  constructed using two different methods.   The first uses
the mean  galaxy number  density at redshift  $z$ calculated  from the
2dFGRS LF.  The second randomizes the coordinates of all mock galaxies
within  the  simulation  box.   Both methods  yield  indistinguishable
estimates of $\xi(r_p,\pi)$ and in what follows we only use the former.
Following H03  each galaxy in a  pair with redshift  separation $s$ is
weighted by the factor
\begin{equation}
\label{Jthree}
w_i = {1 \over 1 + 4 \pi n(z_i) J_3(s)}
\end{equation}
with $n(z)$  the number density  distribution as function  of redshift
and  $J_3(s)  = \int_{0}^{s}  \xi(s')  s'^2  {\rm  d}s'$.  Hence  each
galaxy-galaxy, random-random, and galaxy-random pair is given a weight
$w_iw_j$.  We  substitute $\xi(s')$  with a power  law using  the same
parameters  as in H03.   This redshift  dependent weighting  scheme is
designed  to  minimize  the  variance  on  the  estimated  correlation
function (Davis \& Huchra 1982; Hamilton 1993).
\begin{figure*}
\centerline{\psfig{figure=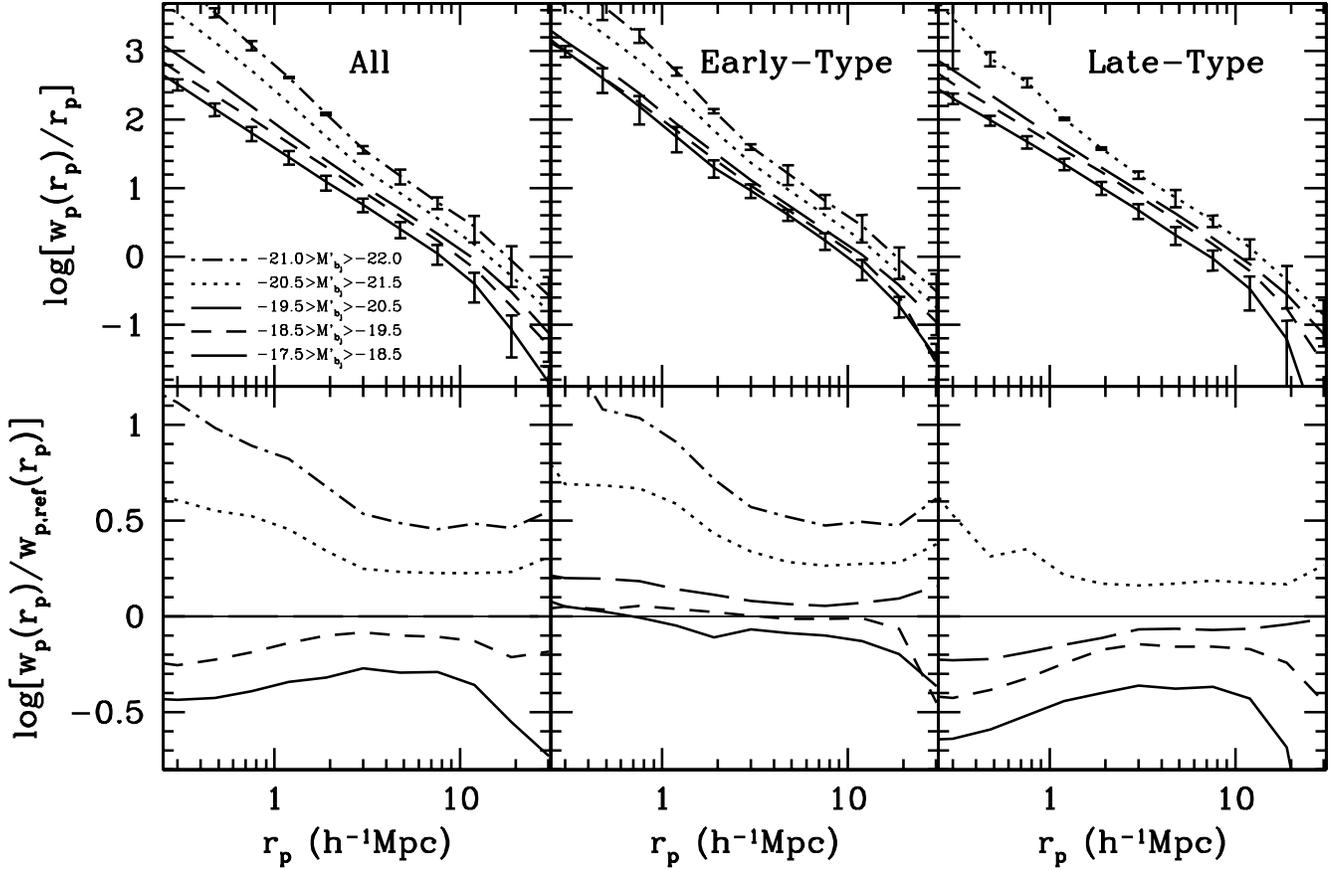,width=\hdsize}}
\caption{The upper panels show  the projected 2PCFs $w_p(r_p)/r_p$ for
galaxies of different luminosity and type. The errorbars correspond to
the 1-$\sigma$  variance among distinct  MGRSs (i.e. among the  8 MSBs
for the  faintest subsamples,  and among the  4 MBs for  the brightest
subsamples).  For clarity,  the error  bars are  only plotted  for the
brightest and  faintest subsamples.  The lower panels  plot the ratios
of  these $w_p(r_p)$  to that  of a  reference sample.   The reference
sample  contains all  galaxies  within the  magnitude  range $-19.5  >
M'_{b_J} > -20.5$ (with $M'_{b_J}  = M_{b_J}-5\log h$).  Note that the
faintest subsamples, which  are impacted by the boxsize  effect of the
$L_{100}$  simulation,  reveal a  `break'  at  $r_p  \simeq 10  h^{-1}
\Mpc$.}
\label{fig:ratio}
\end{figure*}

Since the redshift-space distortions only affect $\pi$, the projection
of  $\xi(r_p,\pi)$  along  the  $\pi$   axis  can  get  rid  of  these
distortions and  give a function that  is more closely  related to the
real-space  correlation function.   In  fact, this  projected 2PCF  is
related to the real-space 2PCF through a simple Abel transform
\begin{equation}
\label{abel}
w_p(r_p) = \int_{-\infty}^{\infty} \xi(r_p,\pi) {\rm d}\pi
= 2 \int_{r_p}^{\infty} \xi(r) \,
{r \, {\rm d}r \over \sqrt{r^2 - r_p^2}}
\end{equation}
(Davis  \& Peebles  1983).  Therefore,  if the  real-space  2PCF is  a
power-law, $\xi(r) = (r_0/r)^\gamma$, the projected  2PCF $w(r_p)$ can
be written as
\begin{equation}
\label{wxipred}
w_p(r_p) = \sqrt{\pi} \, \frac{\Gamma(\gamma/2-1/2)}{\Gamma(\gamma/2)}
\, \left( {r_0 \over r_p} \right)^{\gamma} \, r_p \,.
\end{equation}

We start our investigation of the redshift-space clustering properties
by computing $w_p(r_p)$ for a number of luminosity bins and for early-
and  late-type  galaxies   separately.   To  compare  these  projected
correlation  functions  with the  2dFGRS  results  from Norberg  \etal
(2002a), we  estimate $w_p(r_p)$ using volume-limited  samples with the
same  redshift and magnitude  selection criteria  as those  adopted by
Norberg \etal  (2002a).  For  the MSB  mocks (which use  a stack  of $6
\times 6 \times 6$  $L_{100}$ boxes), however, these $w_p(r_p)$ reveal
a systematic `break'  at $r_p \sim 10 h^{-1} \Mpc$.   As we have shown
in Section~\ref{sec:2pcf}, this owes to  the fact that, because of the
small box-size of  the $L_{100}$ simulation, the 2PCF  is too small on
large  scales (see  Figs~\ref{fig:2pcfdm}  and~\ref{fig:r0}).  We  can
circumvent this problem  by using MGRSs from the the  MB set, in which
the stack  of $6 \times 6 \times  6$ $L_{100}$ boxes is  replaced by a
stack of $2 \times 2 \times  2$ $L_{300}$ boxes.
 However, these  MGRSs are only complete down to
$M_{b_J} - 5 {\rm log} h  \simeq -18.4$ and can therefore only be used
for galaxies brighter than this. 

The upper panels of Fig.~\ref{fig:ratio} plot $w_p(r_p)$ for different
magnitude  bins  and for  early-  and  late-type galaxies  separately.
Except  for the  faintest magnitude  bin, these  projected correlation
functions  are obtained  from MGRS  in the  MB set.   Results  for the
magnitude bin  with $-17.5 > M_{b_J} -  5 {\rm log} h  > -18.5$ (solid
lines) are obtained  from the MSB set.  As  discussed in Paper~II, the
projection  significantly washes  out the  features in  the real-space
2PCFs  at $\sim  2\mpch$, and  the projected  2PCFs better  resemble a
power-law. The exception is  the $w_p(r_p)$ for the faintest subsample
of galaxies, where the `break' mentioned above is clearly visible.  To
highlight the luminosity and  type dependence of $w_p(r_p)$, the lower
panels of  Fig.~\ref{fig:ratio} plot the ratios of  $w_p(r_p)$ to that
of  a reference  sample  defined as  all  (early-type plus  late-type)
galaxies with  $-19.5 > M_{b_J}-5 {\rm  log} h > -20.5$.   For a given
luminosity,  the correlation  amplitude is  higher, and  the  slope is
steeper,  for   early-type  galaxies  than   for  late-type  galaxies.
Significant changes in  the slope (and thus deviations  from a perfect
power-law) occur at separations $r_p  \sim 2 \mpch$, which is at least
qualitatively in  agreement with recent results from  the SDSS (Zehavi
\etal 2003).
\begin{figure*}
\centerline{\psfig{figure=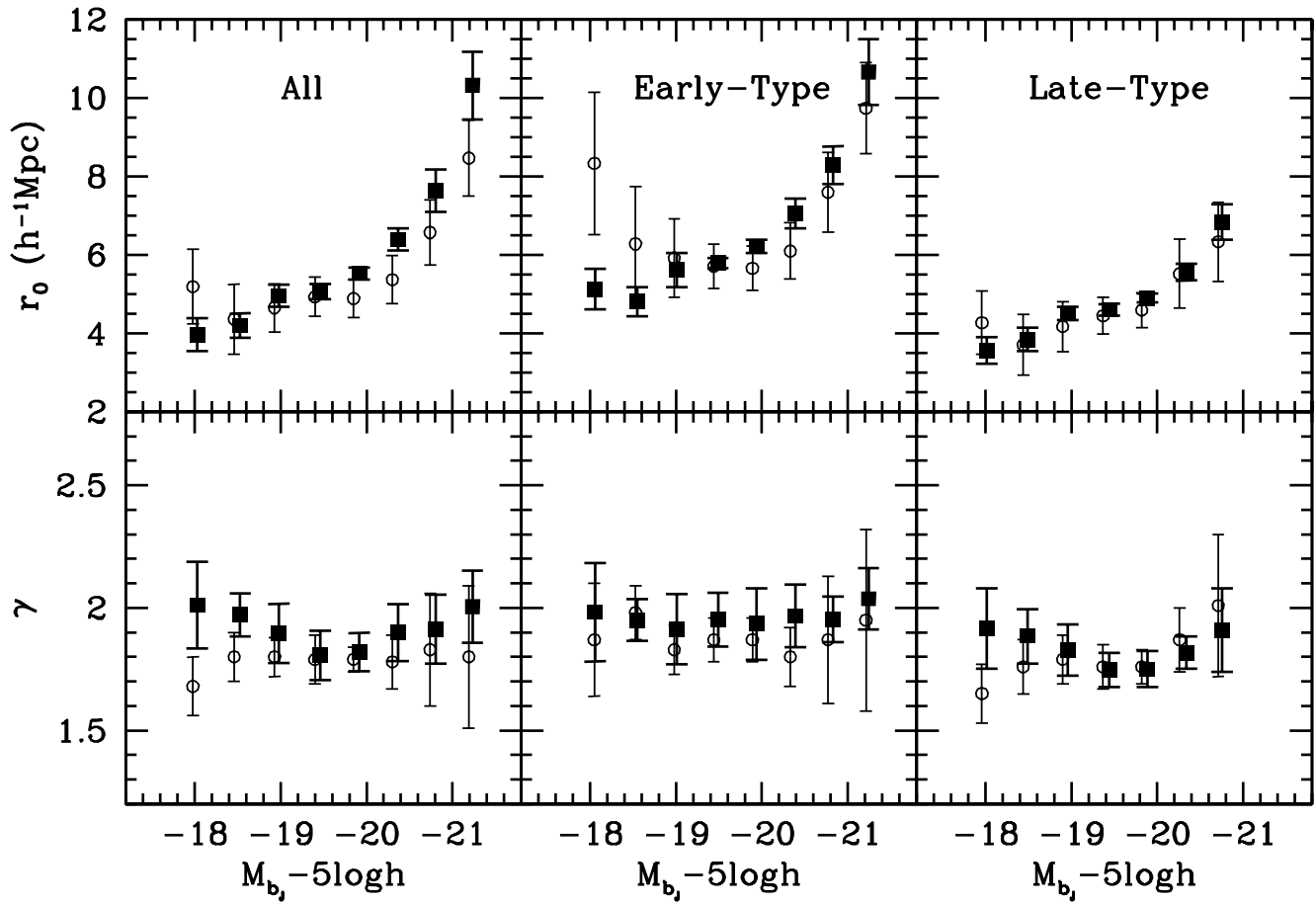,width=\hdsize}}
\caption{The correlation lengths, $r_0$,  and slopes, $\gamma$, of the
power-laws that best fit  the projected correlation functions over the
range $2 \leq r_p \leq 15 \mpch$ (solid squares).  The results for the
2 faintest luminosity  bins are based on the mean  and variance of the
sample of 8  MSB mocks, while results for the other  bins are based on
the mean and  variance of the sample of 4 MB  mocks. Open circles with
errorbars correspond to  the 2dFGRS data of Norberg  \etal (2002a), and
are shown for comparison. Except  for a systematic overestimate of the
correlation  lengths,  the  cause  of  which  has  been  discussed  in
Section~\ref{sec:2pcf}, there is good  agreement between our MGRSs and
the 2dFGRS.}
\label{fig:gama}
\end{figure*}

In order to facilitate a  more direct comparison with the 2dFGRS data,
we  fit a  single power-law  relation of  the  form~(\ref{wxipred}) to
these $w_p(r_p)$ over the range $2 \mpch < r_p < 15 \mpch$. This range
is also  adopted by  Norberg \etal (2002a)  when fitting  the projected
2PCFs  obtained  from   the  2dFGRS.   Fig.~\ref{fig:gama}  plots  the
real-space  correlation lengths  $r_0$  and the  slopes $\gamma$  thus
obtained  as function of  luminosity and  galaxy type.   The agreement
between  our  MGRSs and  the  2dFGRS  is  acceptable. The  slight  but
systematic overestimate  of $r_0$ is  due to the effects  discussed in
Section~\ref{sec:2pcf}.

\begin{figure*}
\centerline{\psfig{figure=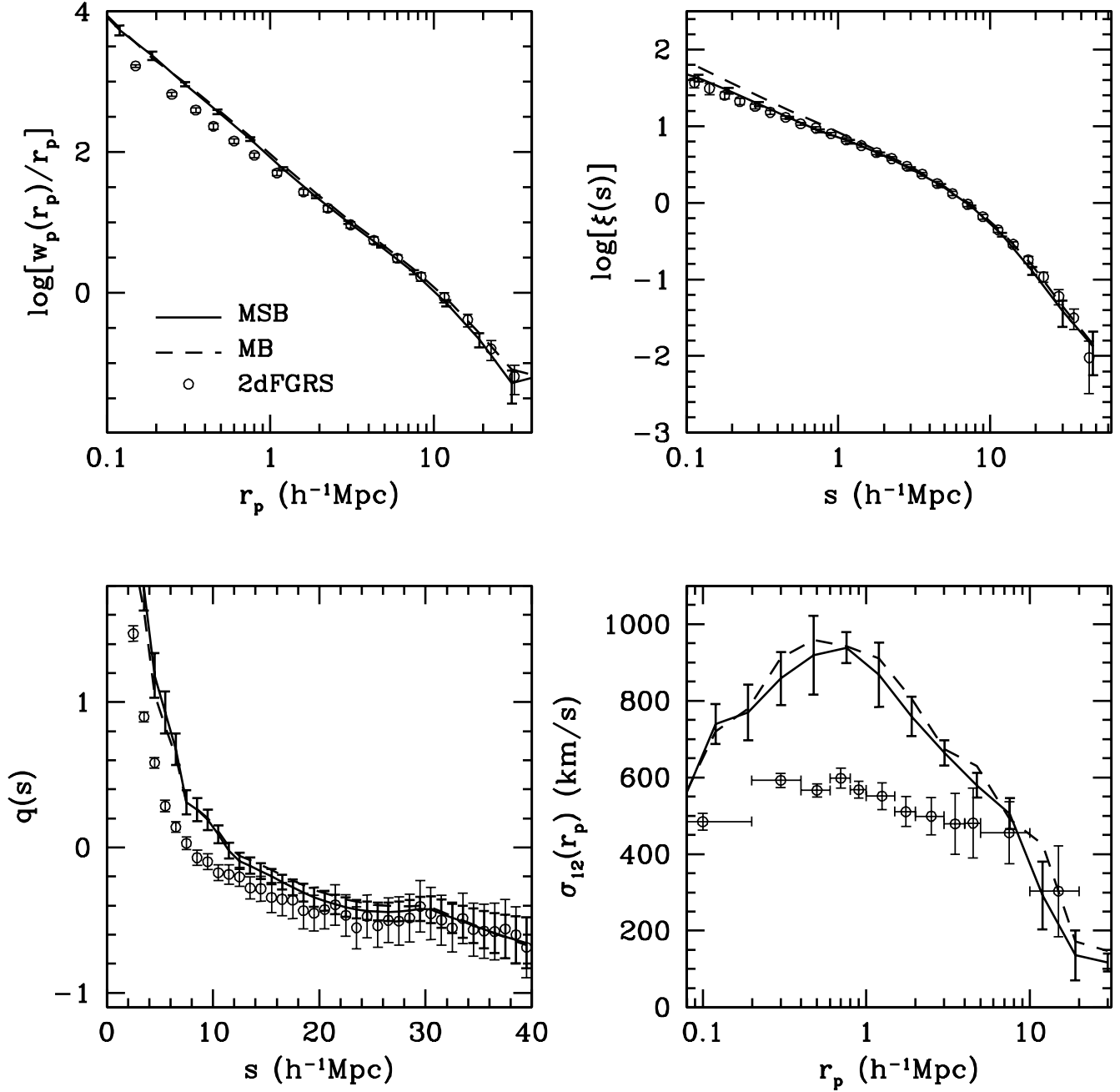,width=\hdsize}}
\caption{The  projected   correlation  function  $w_p(r_p)$  (top-left
panel), the redshift-space  correlation function $\xi(s)$ (top-right),
the  quadrupole-to-monopole ratio $q(s)$  (bottom-left), and  the PVDs
(bottom-right) for  the samples  of MSB (solid  lines) and  MB (dashed
lines) surveys.  Error bars, which are similar for MB and MSB results,
are only  shown for the MSB  results for clarity.  These errorbars are
based on  the variance  of the  8 MSB surveys.  The open  circles with
errorbars correspond  to the 2dFGRS results obtained  by Hawkins \etal
(2003), and are shown for comparison.  Note that the MSBs and MBs give
approximately the same results,  but that there are marked differences
between model predictions and  observations.  Note also that the model
errorbars  are in  general  larger  than the  difference  in the  mean
between  MB  and  MSB  results,  implying  that  these  errorbars  are
statistical.}
\label{fig:resA}
\end{figure*}

We now turn to a  comparison of the projected correlation function for
the   entire,  flux   limited  surveys.    The  upper-left   panel  of
Fig.~\ref{fig:resA} compares  the $w_p(r_p)$  obtained from our  8 MSB
and 4 MB MGRSs with that of the 2dFGRS obtained by H03.  The projected
correlation functions from our MSBs and MBs agree well with each other
(i.e., the 1-$\sigma$  errorbars overlap), and, at $r_p  \gta 3 h^{-1}
\Mpc$, with the 2dFGRS results. Note that at $r_p \gta 10 h^{-1} \Mpc$
the $w_p(r_p)$ obtained from the MB mocks is slightly larger than that
obtained from  the MSB  mocks, again due  to the effects  discussed in
Section~\ref{sec:2pcf}.

At  large scales, $w_p(r_p)$  is predominantly  sensitive to  the halo
occupation numbers $\langle N(M) \rangle$ and virtually independent of
the  second moment  of  $P(N \vert  M)$  or of  details regarding  the
spatial  distribution of  satellite galaxies.   The good  agreement at
large  scales  among  different   MGRSs  and  with  the  observations,
therefore  strongly supports  our CLF  and it  shows that  any `cosmic
variance' among the different MGRSs has only a relatively small impact
on  $w_p(r_p)$.   On small  scales,  however,  the  MGRSs reveal  more
correlation  power (by  about  a  factor 2)  than  observed.  On  such
scales, $w_p(r_p)$  is sensitive to  our assumptions about  the second
moment  of  $P(N  \vert M)$  and,  to  a  lesser degree,  the  spatial
distribution  of   satellite  galaxies.   We  shall   return  to  this
small-scale  mismatch and  its implications  in Section~\ref{sec:disc}
below.

Rather   than  projecting   $\xi(r_p,\pi)$,  one   may   also  average
$\xi(r_p,\pi)$ along constant $s = \sqrt{r_p^2 + \pi^2}$, yielding the
redshift-space   2PCFs    $\xi(s)$.    The   upper-right    panel   of
Fig.~\ref{fig:resA} plots  $\xi(s)$ obtained from  our MGRSs, compared
to the  2dFGRS results from  H03. We find  a similar  behavior as
with the  projected correlation function; the  8 MSBs and  4 MBs agree
quite well  with each  other and  with the observations  at $s  \gta 6
h^{-1}  \Mpc$.  At  smaller redshift-space  separations,  however, the
MGRSs slightly overpredict the correlation power. 
Note that the MB samples  predict higher $\xi(s)$ on small scales than
the  MSB samples.  This  difference comes  from the  fact that  the MB
samples  are incomplete  for  galaxies fainter  than $M_{b_J}-5\log  h
=-18.4$.  To test this we construct a mock survey from the MSB sample,
but only accepting galaxies brighter than this. This yields a $\xi(s)$
in excellent  agreement with that of  the MB samples over  all scales. 
Thus, although  the use of  only large-box simulations can  results in
systematic errors on small scale,  the use of small-box simulations in
the MSB samples  does not cause any signicifant,  systematic errors on
large scale.
\begin{figure*}
\centerline{\psfig{figure=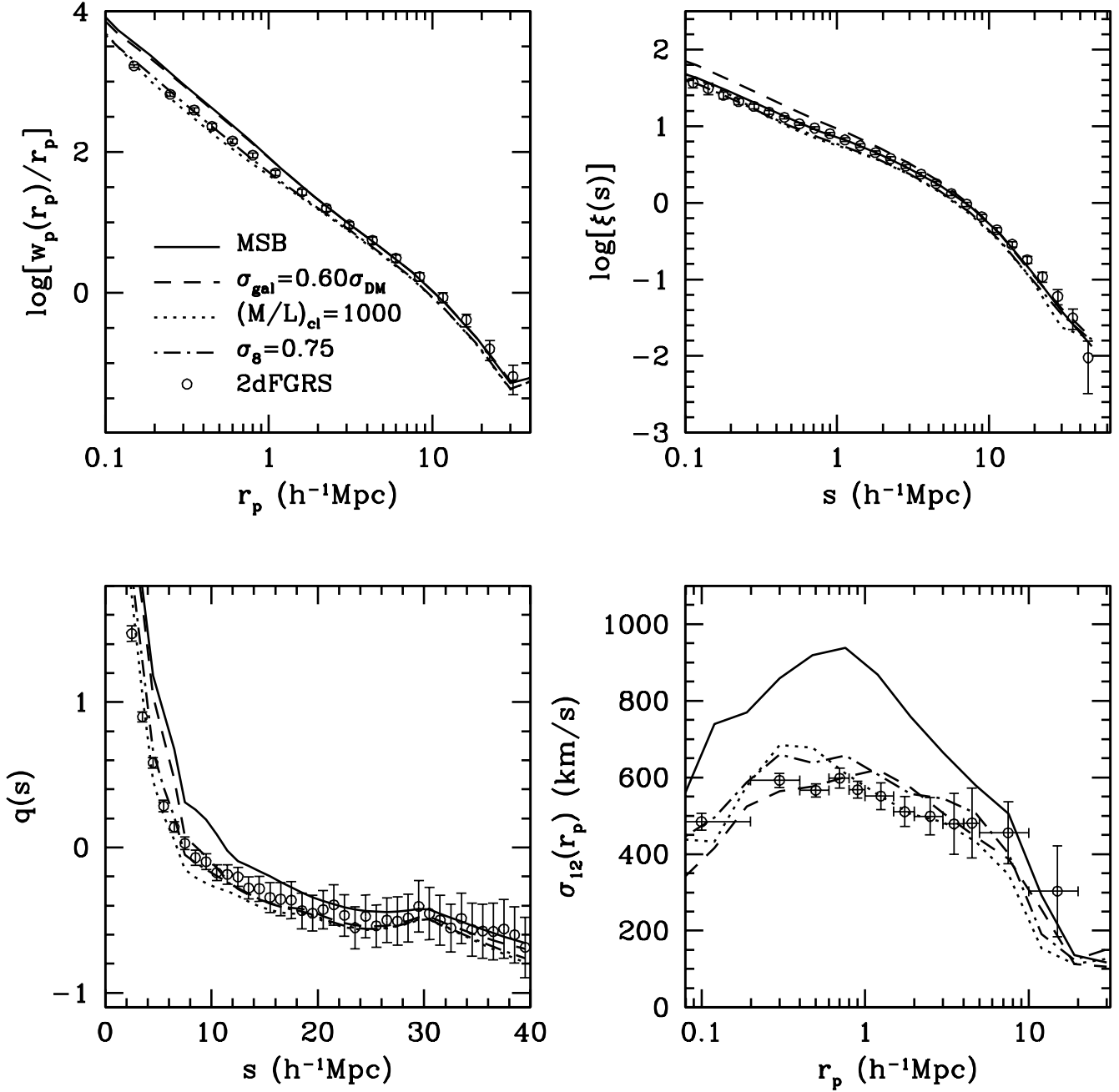,width=\hdsize}}
\caption{Same as  Fig.~\ref{fig:resA} except that here  we compare the
results for  the MSB sample with  those of three  alternative MGRSs in
which we have  modified the CLF to yield  cluster mass-to-light ratios
of $(M/L)_{\rm cl} = 1000 h  \MLsun$ (dotted lines), in which we adopt
a velocity bias of $b_{\rm vel} = 0.6$ (dashed lines), and in which we
adopt  a  cosmology  with  $\sigma_8=0.75$  (dot-dashed  lines).   All
results correspond  to the  mean of  the entire sample  of 8  MSB mock
surveys.  For  clarity, no  errorbars are plotted  here, but  they are
similar  to those shown  in Fig.~\ref{fig:resA}.   Note that  both the
$(M/L)_{\rm cl}=1000 h \MLsun$ model and the $\sigma_8=0.75$ model are
in good agreement with the observational data.}
\label{fig:resB}
\end{figure*}

\subsection{Redshift Space Distortions}
\label{sec:pvd2}

We now turn  to a comparison of the  detailed shape of $\xi(r_p,\pi)$.
In  particular,  we focus  on  the  distortions  with respect  to  the
real-space  correlation  function  $\xi(r)$  induced by  the  peculiar
velocities of galaxies.

The two-dimensional   correlation function  $\xi(r_p, \pi)$  is  often
modeled as  a  convolution of the   real-space 2PCF $\xi(r)$   and the
conditional distribution function $f(v_{12}  \vert r)$:
\begin{eqnarray}
\label{xizmodel}
\lefteqn{1  +      \xi(r_p,   \pi)   =  }        \nonumber \\   &    &
 \int_{-\infty}^{\infty}   \left[    1       +    \xi(\sqrt{r_p^2+(\pi
 -v_{12}/H_0)^2}) \right] \, f(v_{12} \vert r) \, {\rm d}v_{12}
\end{eqnarray}
(Peebles 1980).   Here $v_{12}$  corresponds to the  pairwise peculiar
velocity  {\it along  the line-of-sight}  and $r$  corresponds  to the
real-space  separation.    It  is  standard  practice   to  assume  an
exponential form for $f(v_{12} \vert  r)$ and to ignore its dependence
on separation  $r$ (cf., Davis \&  Peebles 1983; Mo,  Jing \& B\"orner
1993, 1997;  Fisher \etal 1994;  Marzke \etal 1995; Guzzo  \etal 1997;
Jing, B\"orner \& Suto 2002;  Zehavi \etal 2002).  However, as we have
shown  in  Section~\ref{sec:resreal},  the  exponential form  is  only
adequate at small separations, and  the PVD varies quite strongly with
separation.  Furthermore, equation~(\ref{xizmodel})  is only valid for
an isotropic  velocity field in the  limit where the  probability of a
real-space pair separation $r$ is independent of the probability of an
associated relative velocity  $v_{12}$.  Although perhaps a reasonable
approximation on small, highly non-linear, scales, it is certainly not
valid  in linear  theory where  the  velocity and  density fields  are
tightly coupled.   In an  attempt to partially  correct for  this, one
often assumes that $f(v_{12})$ is the probability distribution for the
{\it relative} velocity about the mean.  Using the self-similar infall
model, this mean pairwise peculiar velocity, $\langle v_{12} \rangle$,
is modeled as
\begin{equation}
\label{v12mod}
\langle v_{12} \rangle(r) = -H_0 \, F \,
\left({y \over 1 + (r/r_{0})^2}\right)
\end{equation}
(Davis \&  Peebles 1977)  with $y=|\pi-v_{12}/H_0|$ the  separation in
real-space along  the line-of-sight.  $F=0$ corresponds to  a Universe
without  any  flow  other  than  the  Hubble  expansion,  while  $F=1$
corresponds to stable  clustering.  Given the fairly ad  hoc nature of
this model, and  the strong sensitivity to the  uncertain value of $F$
(Davis \& Peebles 1983), great  care is required when interpreting any
results based on this model.

A more  robust model is based  on linear theory  and directly modeling
the infall velocities  around density perturbations.  Following Kaiser
(1987) and  Hamilton (1992)  one  can write  the observed  correlation
function on linear scales as
\begin{equation}
\label{xileg}
\xi_{\rm lin}(r_p,\pi) = \xi_0(s) {\cal P}_0(\mu) +
\xi_2(s) {\cal P}_2(\mu) + \xi_4(s) {\cal P}_4(\mu).
\end{equation}
Here ${\cal P}_l(\mu)$ is  the $l^{th}$ Legendre polynomial, and $\mu$
is  the  cosine  of  the  angle  between  the  line-of-sight  and  the
redshift-space separation ${\bf s}$. According to linear perturbation
theory the angular moments can be written as
\begin{equation}
\label{ang0mom}
\xi_0(s) = \left( 1 + {2 \beta \over 3} + {\beta^2 \over 5} \right) \,
\xi(r) \,,
\end{equation}
\begin{equation}
\label{ang2mom}
\xi_2(s) = \left( {4 \beta \over 3} + {4 \beta^2 \over 7} \right) \,
\left[ \xi(r) - \overline{\xi}(r) \right] \,,
\end{equation}
\begin{equation}
\label{ang4mom}
\xi_4(s) = {8 \beta^2 \over 35} \, \left[ \xi(r) +
{5 \over 2} \overline{\xi}(r) - {7 \over 2} \hat{\xi}(r) \right]  \,,
\end{equation}
with
\begin{equation}
\label{xibarr}
\overline{\xi}(r) = {3 \over r^3} \int_{0}^{r} \xi(r') r'^2 {\rm d}r'\,,
\end{equation}
and
\begin{equation}
\label{xihatr}
\hat{\xi}(r) = {5 \over r^5} \int_{0}^{r} \xi(r') r'^4 {\rm d}r'
\end{equation}
Given  a value for  $\beta$ and  the real-space  correlation function,
which  can   be  obtained   from  $\xi(r_p,\pi)$  via   the  projected
correlation function $w_p(r_p)$, equation~(\ref{xileg}) yields a model
for $\xi(r_p,\pi)$ {\it on linear scales} that takes proper account of
the coupling  between the density  and velocity fields.  To  model the
non-linear virialized  motions of galaxies within  dark matter haloes,
one  convolves  this $\xi_{\rm  lin}(r_p,\pi)$  with the  distribution
function of pairwise peculiar velocities $f(v_{12} \vert r)$.
\begin{eqnarray}
\label{xiprojmod}
\lefteqn{1 + \xi(r_p, \pi) = } \nonumber  \\
 & & \int_{-\infty}^{\infty} \left[ 1 +
\xi_{\rm lin}(r_p,\pi -v_{12}/H_0) \right] \, f(v_{12} \vert r) \,
{\rm d}v_{12}
\end{eqnarray}

Thus, by modeling $\xi (r_p,\pi)$ one can hope to get both an estimate
of  $\beta$ as  well as  information regarding  the  pairwise peculiar
velocity distribution.  We follow  H03, and assume that the real-space
2PCF  is a  pure  power-law, $\xi(r)  =  (r/r_0)^{-\gamma}$, and  that
$f(v_{12}  \vert r)$  is an  exponential  that is  independent of  the
real-space separation $r$:
\begin{equation}
\label{fmod}
f(v_{12} \vert  r) = f(v_{12})  = {1  \over \sqrt{2} \sigma_{12}} {\rm
exp}\left(- {\sqrt{2} \vert v_{12} \vert \over \sigma_{12}}\right)
\end{equation}
Using a  simple $\chi^2$ minimization technique, we  fit these models,
described by  the four  parameters $\beta$, $\sigma_{12}$,  $r_0$, and
$\gamma$,  to the  $\xi  (r_p,\pi)$ in  each  of our 8 MSB and 4 MB
MGRSs. The $\chi^2$ is defined as
\begin{equation}
\label{chi}
\chi^2=\sum \left( {
{\rm log}[1+\xi]_{\rm model} -  {\rm log}[1+\xi]_{\rm data} \over
{\rm log}[1+\xi+\Delta\xi]_{\rm data} -  {\rm
log}[1+\xi-\Delta\xi]_{\rm data}} \right)^2,
\end{equation}
where  the summation  is over  the $\xi(r_p,\pi)$  data grid  with the
restriction  $8 h^{-1}  \Mpc  < s  <  20 h^{-1}  \Mpc$  (see H03)  and
$\Delta\xi(r_p,\pi)$ is the rms of $\xi(r_p,\pi)$ determined from each
of our 8 MSB  MGRSs (or of our 4 MB MGRSs).   The averages (over the 8
or the  4 MGRSs)  of the best-fit  values for  $\beta$, $\sigma_{12}$,
$r_0$, and $\gamma$, along with the variances among different samples,
are  listed  in the  first  two lines  of  Table~1.   These should  be
compared with the values listed  in the last line, which correspond to
the best-fit values obtained from  the 2dFGRS by H03 using exactly the
same  method. As  one can  see, the  best-fit values  for  $\beta$ and
correlation lengths of the MSB and the MB sample agree with each other
and with the 2dFGRS value at  better than the $1 \sigma$ level. On the
other hand, the discrepancies regarding $\gamma$ and $\sigma_{12}$ are
significant, both of which are  significantly higher in our MGRSs than
in the 2dFGRS.

In order to investigate these discrepancies  in more detail we compute
two statistics of  the redshift space  distortions which we compare to
the 2dFGRS. As  above, we take great care   in using exactly the  same
method and assumptions as H03.  Therefore, even if some aspects of the
model are questionable,  this  allows a  meaningful comparison of  our
results with those  obtained  by H03.  First of  all, we   compute the
modified quadrupole-to-monopole ratio
\begin{equation}
\label{qsdef}
q(s) \equiv  {\xi_2(s) \over {3 \over s^3} \int_{0}^{s}
\xi_0(s') \, s'^2 \, {\rm d}s' - \xi_0(s)}.
\end{equation}
where $\xi_l(s)$ is given by
\begin{equation}
\label{angmom}
\xi_l(s) = { 2l + 1 \over 2} \int_{-1}^{1} \xi(r_p,\pi) \,
{\cal P}_l(\mu) \, {\rm d}\mu.
\end{equation}
The lower-left panel of  Fig.~\ref{fig:resA} plots $q(s)$ for MSBs and
MBs together  with the 2dFGRS  results (open circles  with errorbars).
Although  the MGRSs  reveal the  same overall  behavior as  the 2dFGRS
data,  and are  mutually consistent,  they  systematically overpredict
$q(s)$.   On small scales,  where random  peculiar velocities  cause a
rapid increase  of $q(s)$, this indicates that  the virialized motions
in our  mock surveys  are larger than  observed (see also  below).  On
large scales, where $q(s)$ asymptotes to the linear theory value of
\begin{equation}
\label{qsbeta}
q(s) = { -{4 \over 3}\beta - {4 \over 7}\beta^2 \over
{1 + {2 \over 3}\beta + {1 \over 5}\beta^2}}\,,
\end{equation}
this might indicate that the value of $\beta$ inherent to our MGRSs is
too small  compared to  the real Universe.   On the other  hand, Cole,
Fisher  \& Weinberg  (1994) have  shown that  non-linear,  small scale
power can affect $q(s)$ out to fairly large separations. Therefore the
systematic overestimate of $q$ at large $s$ may simply be a reflection
of  the random  peculiar velocities  being too  large, rather  than an
inconsistency regarding the value of $\beta$.

\begin{table}
\caption{Best fit parameters.}
\begin{tabular}{lcccc}
  \hline
Survey & $\beta$ & $r_0$ & $\gamma$ & $\sigma_{12}$ \\
 (1) & (2) & (3) & (4) & (5) \\
\hline\hline
MSBs          & $0.52\pm 0.05$ & $5.78\pm 0.23$ & $1.99\pm 0.03$ & $687\pm 37$ \\
MBs           & $0.56\pm 0.06$ & $5.93\pm 0.27$ & $1.95\pm 0.03$ & $732\pm 33$ \\
$(M/L)_{cl}$  & $0.52\pm 0.08$ & $4.96\pm 0.21$ & $1.88\pm 0.05$ & $487\pm 44$ \\
$b_{\rm vel}$ & $0.47\pm 0.06$ & $5.99\pm 0.19$ & $2.00\pm 0.06$ & $497\pm 47$ \\
$\sigma_8$    & $0.51\pm 0.06$ & $5.19\pm 0.17$ & $1.91\pm 0.05$ & $505\pm 25$ \\
2dFGRS        & $0.49\pm 0.09$ & $5.80\pm 0.25$ & $1.78\pm 0.06$ & $514\pm 31$ \\
\hline
\end{tabular}
\medskip

The  values  of  $\beta$,   $r_0$  (in  $h^{-1}\Mpc$),  $\gamma$,  and
$\sigma_{12}$  (in $\kms$)  that best  fit the  $\xi(r_p,\pi)$  for $8
h^{-1} \Mpc  < s <  20 h^{-1} \Mpc$  for a number of  different MGRSs.
Note that  4 MGRSs are used  for MBs, while  8 MGRSs are used  for all
other cases.  The quoted values are the mean and $1\sigma$ variance of
these MGRSs. The MGRSs denoted by `$(M/L)_{\rm cl}$' is similar to the
MGRSs  in the  MSB set,  except that  here the  CLF is  constrained to
mass-to-light ratios for clusters of $(M/L)_{\rm cl} = 1000 h \MLsun$,
rather  than  $(M/L)_{\rm  cl}  =   500  h  \MLsun$  as  in  MSB  (see
Section~\ref{sec:clml}).  The  MGRSs  denoted  by `$b_{\rm  vel}$'  is
similar  except for  a velocity  bias  of $b_{\rm  vel} =  \sigma_{\rm
gal}/\sigma_{\rm  DM}  =  0.6$ (see  Section~\ref{sec:velbias}).   The
MGRSs denoted by `$\sigma_8$' is  also similar except that it adopts a
flat  $\Lambda$CDM cosmology  with $\sigma_8=0.75$  rather  than $0.9$
(see  Section~\ref{sec:sigma}).  The  final  line lists  the  best-fit
parameters   obtained  by   Hawkins  \etal   (2003)  by   fitting  the
$\xi(r_p,\pi)$  obtained from  the  2dFGRS. Note  that  the errors  in
Hawkins \etal are estimated from the spread of 22 Mock samples.

\end{table}

The  second  statistic that  we  use  to  compare the  redshift  space
distortions  in our  MGRSs  with those  of  the 2dFGRS  are the  PVDs,
$\sigma_{12}(r_p)$,  as   a  function  of   projected  radius,  $r_p$.
Following  H03, we  keep  $r_0$,  $\gamma$ and  $\beta$  fixed at  the
`global' values listed in  Table~1 and determine $\sigma_{12}(r_p)$ by
minimizing    $\chi^2$   in    a   number    of    independent   $r_p$
bins\footnote{Note that the  PVDs thus obtained are a  kind of average
of the true PVD along the line-of-sight.  Therefore, these PVDs should
not   be    compared   directly   to    the   true   PVD    shown   in
Fig.~\ref{fig:pvd}.}.  The results are  shown in the lower-right panel
of    Fig.~\ref{fig:resA}.     Whereas    the   2dFGRS    reveals    a
$\sigma_{12}(r_p)$  that  is  almost  constant with  radius  at  about
$500$--$600   \kms$,  our   MGRSs  reveal   a  strong   increase  from
$\sigma_{12} \sim 600 \kms$ at $r_p = 0.1 h^{-1} \Mpc$ to $\sigma_{12}
\sim 900 \kms$ at $r_p =  1.0 h^{-1} \Mpc$, followed by a decrease to
$\sigma_{12}  \sim 500  \kms$ at  $r_p =  10 h^{-1}  \Mpc$.   Thus, at
around $1  h^{-1} \Mpc$, our MGRSs dramatically  overestimate the PVD.
Although  there  is  a  non-negligible  amount of  scatter  among  the
different mock surveys, reflecting the extreme sensitivity of the PVDs
to the few richest systems in the survey, the variance among the 8 (4)
MGRSs is small compared to the discrepancy.

As shown by Peacock \etal  (2001) the best-fit values of $\sigma_{12}$
and $\beta$ are highly degenerate.   We have tested the impact of this
degeneracy on  our $\sigma_{12}(r_p)$  by repeating the  same exercise
using  a value for  $\beta$ that  is $0.1$  larger (smaller)  than the
values listed  in Table~1.   This leads to  an increase  (decrease) of
$\sigma_{12}(r_p)$ of the order of 5 percent (20 percent) at projected
radii of  $1 h^{-1}  \Mpc$ ($10 h^{-1}  \Mpc$).  Given that  our MGRSs
overpredict the PVD at $r_p=1 h^{-1}  \Mpc$ by about 70 percent, it is
clear   that   this  discrepancy   is   not   a   reflection  of   the
$\beta$--$\sigma_{12}$  degeneracy.  Thus,  the  standard $\Lambda$CDM
model seems to have a severe  problem in matching the observed PVDs on
intermediate scales.

\section{Towards a Self-Consistent Model for Large Scale Structure}
\label{sec:disc}

Our  MGRSs, based on  a flat  $\Lambda$CDM concordance  cosmology with
$\Omega_m=0.3$ and  $\sigma_8=0.9$, and on  a CLF that is  required to
yield cluster  mass-to-light ratios of $(M/L)_{\rm  cl}=500 h \MLsun$,
reveals clustering statistics that are overall in reasonable agreement
with the  data from the 2dFGRS.  Nevertheless,  two discrepancies have
come to  light: the MGRSs predict  too much power on  small scales and
PVDs that are too high.  We now investigate possible ways to alleviate
these discrepancies.

\subsection{Halo occupation models}
\label{sec:hom}

The discrepancies  between   our MGRS and  the   2dFGRS  results might
indicate a problem with our halo  occupation models.  Although the CLF
is fairly well constrained by the observed luminosity function and the
observed  luminosity  dependence  of   the   correlation lengths  (see
Papers~I and~II), we  have made  additional assumptions regarding  the
second moments    of the  halo  occupation number    distributions and
regarding  the distribution of  galaxies within individual dark matter
haloes.

As we have shown  in Section~\ref{sec:2pcf} the real space correlation
function  depends only very  weakly   on our  method  of  distributing
satellite    galaxies           within        dark  matter      haloes
(cf. Fig.~\ref{fig:cgrg}). We have  verified, using a number of tests,
that modifications of  the spatial distribution  of satellite galaxies
within dark matter haloes  have no significant influence on $w_p(r_p)$
or on  $\sigma_{12}(r_p)$.    Therefore, none  of  the   discrepancies
mentioned above can be attributed to errors in our satellite model.

Our results  are more susceptible to  changes in the  second moment of
the  halo occupation number distributions.   On small scales, $\xi(r)$
scales with the  average number of  galaxy pairs in  individual haloes
$\langle {1 \over 2} N (N-1) \rangle$. Therefore, one can decrease the
power on  small scales, to  bring our  $w_p(r_p)$  in better agreement
with  observations,  by   lowering the  second  moments  of  our  halo
occupation  distributions.   However, our distribution~(\ref{pnm})  is
already the narrowest  distribution possible, and modifying the second
moment  of    the halo occupation   distributions  can  therefore only
aggravate the discrepancies on small scales.

\subsection{Velocity Bias}
\label{sec:velbias}

A seemingly obvious   explanation for the too high   PVDs is that  the
peculiar velocities of galaxies are  biased with  respect to the  dark
matter.  We define the   velocity bias (sometimes called   `dynamical'
bias)   as $b_{\rm   vel} = \sigma_{\rm   gal}/\sigma_{\rm DM}$,  with
$\sigma_{\rm gal}$   and   $\sigma_{\rm DM}$  the   peculiar  velocity
dispersions of (satellite)  galaxies and  dark  matter particles in  a
given halo,  respectively.  Note that in our  fiducial  MGRSs we adopt
$b_{\rm vel}=1$ (i.e.   no velocity  bias).  Fig.~\ref{fig:resB} shows
$w_p(r_p)$, $\xi(s)$,  $q(s)$ and $\sigma_{12}(r_p)$  for MGRSs (with
the MSB configuration of simulation boxes) in   which $b_{\rm  vel} =
0.6$;  i.e., the velocity dispersion of  satellite galaxies is only 60
percent of that of the dark matter particles in  the same halo (dashed
lines).  With  such a pronounced  velocity  bias, both  $q(s)$ and the
PVDs, as well as $\beta$ and the  global value of $\sigma_{12}$ listed
in Table~1, are all consistent with the 2dFGRS results.

In two recent papers, Berlind  \etal (2003) and Yoshikawa \etal (2003)
measured  the  velocity bias  of  `galaxies'  in  a smoothed  particle
hydrodynamics (SPH) simulation and  found that $b_{\rm vel}$ decreases
from $b_{\rm  vel} \sim 0.9$--$1.0$ for  haloes with $M  \sim 3 \times
10^{14} h^{-1} \Msun$ to $b_{\rm vel} \sim 0.6$--$0.8$ for haloes with
$M \sim  3 \times 10^{12}  h^{-1} \Msun$.  Thus,  although simulations
predict that low-mass  haloes might have values for  the velocity bias
as low as $b_{\rm vel} \simeq 0.6$, required here to bring our PVDs in
agreement  with observations,  the  PVD is  dominated  by galaxies  in
massive  haloes for  which these  same simulations  apparently predict
close to  $b_{\rm vel}=1$ (i.e., no velocity  bias).  Furthermore, the
introduction of velocity  bias cannot solve the excess  power on small
scales.  After all, the real-space correlation function is independent
of $b_{\rm  vel}$ such that the discrepancies  regarding $w_p(r_p)$ on
small scales remain (see upper-left panel of Fig.~\ref{fig:resB}).  To
make  matters  worse, reducing  the  peculiar  velocities of  galaxies
inside dark  matter haloes, {\it  increases} the small scale  power in
{\it redshift space}.  This means, that the redshift space correlation
function  $\xi(s)$ becomes  actually more  discrepant with  the 2dFGRS
data  (see  upper-right  panel  of  Fig.~\ref{fig:resB}).   Therefore,
although some  amount of  velocity bias might  be expected, we  do not
consider it a viable solution for the problems mentioned above.

\subsection{Cluster mass-to-light ratios}
\label{sec:clml}

Because the PVD is a pair weighted statistic it is extremely sensitive
to the few richest  systems in the sample (i.e.,  Mo, Jing \& B\"orner
1993,  1997; Zurek \etal  1994; Marzke \etal 1995; Somerville, Primack
\& Nolthenius 1997). The fact that the PVDs in our MGRSs are too large
compared with observations therefore might  indicate that either there
are  too many clusters   of   galaxies  in   our mock  surveys    (see
Section~\ref{sec:sigma} below), or  that  these clusters  contain  too
many galaxies.

Our CLF  was   constructed  under the   constraint  that the   average
mass-to-light ratio of  haloes with $M  \geq 10^{14} h^{-1}  \Msun$ is
equal to $(M/L)_{\rm cl}  = 500 h  \MLsun$  (in the photometric  $b_J$
band). This  value is motivated by  the average mass-to-light ratio of
clusters obtained by Fukugita, Hogan \& Peebles (1998).  To reduce the
number  of galaxies per cluster we now  set $(M/L)_{cl}=1000 h \MLsun$
and repeat  the entire exercise: we first  use the method described in
Section~\ref{sec:clf} to compute the parameters of the new conditional
luminosity function.  This CLF is used  to construct new MGRSs (using
the same configuration  of simulation boxes as in MSB), from  which we
determine the same statistics as before.

The  results  are listed  in  Table~1 and  shown  as  dotted lines  in
Fig.~\ref{fig:resB}.  Clearly,  increasing the mass-to-light  ratio of
clusters lowers $q(s)$ and $\sigma_{12}(r_p)$, bringing them in better
agreement  with  the 2dFGRS  results.   Although  the  PVDs are  still
somewhat  too high,  especially at  around $\sim  0.4 h^{-1}  \Mpc$, the
extent of  this discrepancy is  similar to its 1-$\sigma$  variance of
the 8  MGRSs, indicating that this remaining  difference is consistent
with `cosmic variance'.  As can be seen from Table~1, both $\beta$ and
$\gamma$ are now in much better  agreement with the 2dFGRS.  In addition,
the  reduction of  the number  of galaxies  in  clusters significantly
reduces    $w_p(r_p)$   at    small    projected   separations    (see
Fig.~\ref{fig:resB}),   bringing  it  in   good  agreement   with  the
observations
\footnote{Note that the correlation amplitude predicted with 
$(M/L)_{cl}=1000 h \MLsun$ is slightly lower than the observed 
amplitude, because in this model more galaxies are assigned to 
small haloes in order to match the observed luminosity function.
Since the errorbars on the observed correlation lengths
are relatively large, the model tends to compromise
the accuracy of the fit to the correlation lengths.}.  
A  similar reduction  of  small  scale  power is  also
evident  in $\xi(s)$.   Thus, these  particular MGRSs  have clustering
characteristics  that are overall  in good  agreement with  the 2dFGRS
results.   The  question is  therefore  whether  or  not such  a  high
mass-to-light  ratio  for  clusters  of galaxies  is  compatible  with
observations.

The cluster  mass-to-light ratios quoted by Fukugita  \etal (1998) are
$(450\pm 100)h({\rm  M/L})_\odot$ in the  $B$-band based on  X-ray and
velocity-dispersion  data.   Taking these  numbers  at  face value,  a
cluster mass-to-light ratio of $(M/L)_{\rm cl}=1000 h \MLsun$ is ruled
out at  the $5\sigma$  level. Using a  variety of methods  to estimate
cluster  masses, Bahcall \etal  (2000) obtained  $(M/L)_B=(330\pm 77)h
({\rm M/L})_\odot$,  which is consistent with the  results of Carlberg
\etal (1996), $(M/L)_B=(363\pm 65)h ({\rm M/L})_\odot$ based on galaxy
kinematics in clusters.  Taking  the average of these two measurements
yields $\langle M/L_B \rangle_{\rm cl} = (350 \pm 70) h \MLsun$, which
rules  out  the cluster  mass-to-light  ratio  required  to match  the
clustering  power on  small scales  at  more than  $7 \sigma$.   Thus,
unless  the   cluster  mass-to-light  ratios   obtained  from  current
observations are  seriously in  error, increasing the  average cluster
mass-to-light ratio to  $(M/L)_{\rm cl} \simeq 1000 h \MLsun$ does not
seem a viable solution for the problems at hand.

\subsection{Power spectrum normalization}
\label{sec:sigma}

Rather than  lowering the average  number of galaxies per  cluster, we
may  also hope  to lower  the PVDs  by reducing  the actual  number of
clusters. As we have shown in Section~\ref{sec:pvd2}, our results, and
thus the  number density  of rich clusters,  is robust  against cosmic
variance.  Therefore, a  lower number  density of  clusters  implies a
different cosmological model.  It is well known that  the abundance of
(rich)   clusters  is  extremely   sensitive  to   the  power-spectrum
normalization parameter  $\sigma_8$.  The too high PVDs  could thus be
indicative of a too high value for $\sigma_8$.

We therefore wish to compute the PVDs in a $\Lambda$CDM cosmology with
identical   cosmological    parameters   as   before,    except   that
$\sigma_8=0.75$ rather than $0.9$. 
Note that the choice of
$\sigma_8=0.75$ is somewhat arbitrary, but it does represent
a compromise between the constraints on the value of $\sigma_8$ 
from various observations and the low value required by our 
results on the PVDs (see below). 
In principle, constructing new MGRSs
for  a different cosmology  requires new  $N$-body simulations  of the
dark  matter  distribution.   This,  however, is  computationally  too
expensive, which is  why we use an approximate  method instead. First,
we compute the new best-fit  parameters of the CLF for this cosmology,
again demanding that $(M/L)_{\rm cl} = 500 h \MLsun$. Next we populate
the  dark matter haloes  in our  $\sigma_8=0.9$ simulation  boxes with
galaxies  according to  this new  CLF.   Finally, we  construct a  new
sample of 8 MSB MGRSs, in which we weigh each galaxy in a halo of mass
$M$ by
\begin{equation}
\label{cosmoweight}
w = {n(M \vert \sigma_8=0.75) \over n(M \vert \sigma_8=0.9)}
\end{equation}
with  $n(M)$ the number  density of  dark matter  haloes of  mass $M$.
This, to first order, mimics  the effect of lowering $\sigma_8$ on the
halo mass  function, and  so should be  a reasonable  approximation on
small  scales where the  clustering properties  are determined  by the
galaxy distribution in individual haloes\footnote{Note that for haloes
with a  given mass,  the concentration parameters  are smaller  in the
$\sigma_8=0.75$  model  than they  are  in  the standard  $\Lambda$CDM
model.   This change  of concentration  is taken  into account  in our
analyses, eventhough  its effect is almost  negligible.}.  The results
for these MGRSs are  shown as dot-dashed lines in Fig.~\ref{fig:resB}.
As one  can see, the agreement with observational results in this
model is much better  than in the standard $\Lambda$CDM model.
These  results suggest  that a  $\Lambda$CDM model  with $\sigma_8\sim
0.75$  may  match all  the  observational  results  obtained from  the
2dFGRS. Unfortunately, in the absence of proper $N$-body simulations 
for this model, it is impossible to make a more detailed comparison 
with observation.

The  question  is,  of  course,  whether  such  a  low  $\sigma_8$  is
compatible with other  independent observations.  Currently, the value
of $\sigma_8$  is constrained mainly  by three types  of observations:
weak lensing  surveys, cluster mass  functions, and anisotropy  in the
cosmic  microwave background. Recent  cluster abundance  analyses give
values of  $\sigma_8$ (assuming $\Omega_m=0.3$) in a  wide range, from
$0.6$ to $1$  (e.g. Borgani \etal 2001; Seljak  2002; Viana, Nichol \&
Liddle 2002;  Pen 1998; Fan  \& Bahcall 1998; Reiprich  \& B\"ohringer
2002). Results  from weak lensing surveys are  equally uncertain, with
$\sigma_8$ spanning  the range $0.7$  to $\sim 1$ (e.g.,  Jarvis \etal
2003;  Hoekstra, Yee  \& Gladders  2002; Refregier  \etal  2002; Bacon
\etal   2003).   Thus,  our   preferred  value,   $\sigma_8=0.75$,  is
consistent with these observations.  At the moment, the most stringent
constraint  on the  value of  $\sigma_8$ is  from WMAP  (Spergel \etal
2003): $\sigma_8=0.84  \pm 0.04$ ($1\sigma$ error).   Even taking this
result at  face value,  one cannot rule  out $\sigma_8=0.75$  with any
high confidence.   Thus, there is no strong  observational evidence to
argue against a $\Lambda$CDM model with $\sigma_8=0.75$.  Furthermore,
as  discussed  in  van den  Bosch,  Mo  \&  Yang  (2003), a  value  of
$\sigma_8$  as  low as  $0.75$  can  also  help to  alleviate  several
problems in current models of  galaxies formation, such as the ones in
connection  to the  Tully-Fisher relation  and to  the  rotation curve
shapes of low-surface brightness galaxies.  Our results presented here
give additional support for a relatively low value of $\sigma_8$.

\section{Conclusions}
\label{sec:concl}

In this  paper, we have used realistic  halo occupation distributions,
obtained   using  the   conditional   luminosity  function   technique
introduced  by Yang,  Mo \&  van den  Bosch (2003),  to  populate dark
matter  haloes  in  high-resolution  simulations of  the  $\Lambda$CDM
`concordance'  cosmology.   The simulations  follow  the evolution  of
$512^3$ dark matter  particles in periodic boxes of  $100 h^{-1} \Mpc$
and $300 h^{-1} \Mpc$ on a side.  Subsequently, the dark matter haloes
identified  in  these  simulations  are  populated  with  galaxies  of
different luminosity and different morphological type.

We  have  shown that  the  luminosity  functions  and the  correlation
lengths  as  function of  luminosity,  both  for  the early-  and  the
late-type galaxies,  are in  good agreement with  observations.  Since
these  same  observations  were  used  to  constrain  the  conditional
luminosity functions,  which in  turn were used  to populate  the dark
matter  haloes,   this  agreement  shows  that   the  halo  occupation
statistics  obtained  analytically  can  be  implemented  reliably  in
$N$-body  simulations to  construct  realistic, self-consistent,  mock
galaxy  distributions. We have  demonstrated that  the details  of the
spatial distribution of galaxies  within individual dark matter haloes
have only  a very mild  effect on the two-point  correlation function,
and only at real-space separations $r \lta 0.3 h^{-1} \Mpc$.

The  mean  pairwise  peculiar  velocities, $\langle  v_{12}  \rangle$,
however,  depend rather  strongly on  whether satellite  galaxies (any
galaxy in  a dark  matter halo other  than the most  luminous, central
galaxy)  are  associated with  random  dark  matter  particles of  the
friends-of-friends (FOF) group, or  whether they are assigned peculiar
velocities  assuming  a  spherical,  isotropic  velocity  distribution
around  the  central galaxy.   In  the  former  case, $\langle  v_{12}
\rangle$,  which indicates  the  amount of  infall around  overdensity
regions, is  similar to that of  the dark matter. In  the latter case,
$\langle v_{12}  \rangle$ is significantly suppressed  with respect to
the dark matter. This difference indicates that the outer parts of the
FOF-groups are not yet virialized.

The pairwise velocity dispersions (PVDs)  of the galaxies are found to
be significantly  smaller than those of dark  matter particles.  Since
the PVD  is a pair  weighted measure for  the potential well  in which
dark  matter particles (galaxies)  reside, this  can be  understood as
long  as the average  number of  galaxies per  unit halo  mass, $N/M$,
decreases  with $M$ (Jing  \etal 1998).   Indeed, the  halo occupation
numbers  inferred from  our conditional  luminosity  function indicate
that $N/M \propto M^{a}$ with $a \sim -0.2$.

Stacking  a  number  of  $100  h^{-1}  \Mpc$  and  $300  h^{-1}  \Mpc$
simulation boxes  allows us to construct mock  galaxy redshift surveys
(MGRSs) that  are comparable to the  2dFGRS in terms  of sky coverage,
depth, and magnitude  limit.  For each of these  MGRSs we estimate the
two-point  correlation functions  $\xi(r_p,\pi)$.  These  are  used to
derive a  number of statistics  about the large scale  distribution of
galaxies,  which we  compare  directly with  the  2dFGRS results.   In
particular, we calculate the projected 2PCFs $w_p(r_p)$ as function of
luminosity  and type.   The best-fit  power-law slope  and correlation
lengths  of these projected  correlation functions  are found  in good
agreement with the 2dFGRS results obtained by Norberg \etal (2002a). In
addition,  we   also  compute   $w_p(r_p)$  and  the   redshift  space
correlation  function  $\xi(s)$  for  the  entire  MGRSs.   These  are
compared   to   the  2dFGRS   results   obtained   by  Hawkins   \etal
(2003).

Although the  agreement with  the 2dFGRS data  is excellent  on scales
larger than $\sim 3 h^{-1}\Mpc$, on smaller scales $w_p(r_p)$
is about a factor  two larger than observed.  To investigate
this  in  more  detail,  we analyzed  the  redshift-space  distortions
present  in  $\xi(r_p,\pi)$  by computing  the  quadrupole-to-monopole
ratios    $q(s)$    and     the    pairwise    velocity    dispersions
$\sigma_{12}(r_p)$.  A  comparison with  the results of  Hawkins \etal
(2003) shows  that the standard  $\Lambda$CDM model  over-predicts the
clustering power  on small scales  by a factor  of about two,  and the
PVDs by about $350 \kms$.  After examining a variety of possibilities,
we find that  the only viable solution to these  problems is to reduce
the power spectrum amplitude, $\sigma_8$, from $0.9$ to $\sim 0.75$.

No doubt,  in the coming  years, new results  from the 2dFGRS  and the
SDSS  will   significantly  improve  the  data  on   the  large  scale
distribution of  galaxies. The analysis  presented here, based  on the
conditional luminosity function, will hopefully prove a useful tool to
further  constrain  both  galaxy  formation and  cosmology.   In  this
respect,  our  results  regarding  constraints on  $\sigma_8$  are  an
important illustration of the potential power of this approach.


\section*{Acknowledgement}

We thank  Ed Hawkins and the  2dFGRS team for providing  us with their
data  in  electronic  format,  the anonymous  referee  for  insightful
comments  that helped  to  improve the  paper,  and Gerhard  B\"orner,
Guinevere  Kauffmann,  Simon  White  and  Saleem  Zaroubi  for  useful
discussions.   XHY   is  supported   by  NSFC  (No.10243005)   and  by
USTCQN. YPJ  is supported  in part by  NKBRSF (G19990754) and  by NSFC
(No.10125314).  Numerical  simulations presented  in  this paper  were
carried out  at the  Astronomical Data Analysis  Center (ADAC)  of the
National Astronomical Observatory, Japan.


\label{lastpage}

\end{document}